\documentstyle[11pt,aaspp4]{article}

\begin{document}

\title{Virialization of Galaxy Clusters and Beyond}

\author{Wen Xu\footnote{Department of Physics and Astronomy, Arizona State
University, Tempe, AZ 85287;
and Beijing Astronomical Observatory,
Chinese Academy of Sciences, Beijing 100012, China},
Li-Zhi Fang\footnote{Department of Physics, University of Arizona,
Tucson, AZ 85721} and
Xiang-Ping Wu\footnote{Beijing Astronomical Observatory, 
Chinese Academy of Sciences, Beijing 100012, China}
}

\begin{abstract}

Using samples of structures identified by a multi-scale decomposition
from numerical simulation, we analyze the scale-dependence of the
virialization of clusters. We find that beyond the scale of full
virialization there exists a radius range over which clusters are
quasi-virialized, i.e. while the internal structure of an {\it individual}
cluster is at substantial departure from dynamical relaxation, some
{\it statistical} properties of the multi-scale identified clusters
are approximately the same as those for the virialized systems. 
The dynamical reason of the existence of quasi-virialization is that
some of the scaling properties of dynamically relaxed systems of cosmic
gravitational clustering approximately hold beyond the full
virialization regime. This scaling can also be seen from a semi-analytic
calculation of the mass functions of collapsed and
uncollapsed halos in the Press-Schechter formalism.

The ``individual-statistical" duality of the quasi-virialization provides
an explanation of the observed puzzle that the total masses of clusters
derived from virial theorem are statistically the same as the gravitational
lensing determined masses, in spite of the presence of irregular
configuration and substructures in individual clusters. It also explains
the tight correlation between the velocity dispersion of optical galaxies
and the temperature of X-ray emitting gas. Consequently, the virial mass
estimators based on the assumptions of isothermal and hydrostatic model
are statistically applicable to scales on which the clusters are
quasi-virialized. In the quasi-virialization regime, the temperature
functions of clusters also show scaling. This feature is a useful
discriminator among models. As a preliminary comparison with observation, the
discriminator yields a favor of the models of LCDM and OCDM.

\end{abstract}
\vspace{-5mm}
\keywords{cosmology: theory --- galaxies: clustering ---
large scale structure of universe --- methods: numerical}

\section{Introduction}

Clusters of galaxies are playing an important role in the study of large
scale structures of the universe. In particular, the observed dynamical
properties of clusters set stringent constrains on models of structure
formation as well as on cosmological parameters. The popular inflation
plus the cold dark matter (CDM) cosmologies are found to be able to
match the existing observations of clusters if the cosmological
parameters such as the mass density $\Omega_0$, the cosmological constant
$\Omega_{\Lambda}$ and the mass fluctuation amplitude are properly selected
(e.g. Bahcall \& Cen 1992; Jing et al. 1993, 1995; Jing \& Fang 1994;
Eke et al. 1996; Viana \& Liddle 1996; Carlberg et al. 1997; Bahcall,
Fan \& Cen 1997; Xu, Fang \& Wu 1998; Xu, Fang \& Deng 1999).

Yet, some fundamental properties of clusters are not directly measurable,
among which the conventional determination of gravitating masses of clusters 
is related to the dynamical state of the clusters, i.e. the virialization of
both baryonic and nonbaryonic matter. For instance, the so-called
dynamical mass estimator relies on the hypothesis that the optical galaxies
and/or the hot intracluster gas are the tracers of the underlying
gravitational potential. Nevertheless, the virialization assumption is
often challenged by the existence of substructures
(e.g. Richstone, Loeb \& Turner 1992; Jing et al. 1995;
Crone, Evrard \& Richstone 1996; Patrick 1999;
Mathiesen, Evrard \& Mohr 1999).
Substructures in both optical and X-ray
maps of clusters suggest that clusters may still be in the
process of formation (e.g. Stein 1997; Kriessler \& Beers 1997;
Solanes et al. 1999).
Spatially-resolved measurements of gas temperature in some clusters
illustrate the complex two-dimensional patterns including the asymmetric
variation and the significant decline with radius (Henry \& Briel 1995;
Henriksen \& White 1996; Henriksen \& Markevitch 1996; Markevitch 1996),
indicative of strong substructure merging. The unidentified or neglected
substructures in clusters may lead to an overestimate of galaxy velocity
dispersion, and therefore the virial mass (Smail et al. 1997). This
implies that the hydrostatic equilibrium hypothesis may be inappropriate to
the dynamical state of optical galaxies and intracluster gas for some
clusters.

On the other hand, a statistical comparison of the cluster masses determined
from optical galaxies, X-ray observations and gravitational lensing shows
that there is an excellent agreement among different mass estimates on
scales much greater than the core radii. The discrepancy of these mass
estimators appears only inside the central regions of clusters where the
local dynamical activities of galaxies and cooling flows may become
dominant (Wu \& Fang 1997; Allen 1998; Wu et al. 1998). Since gravitational
lensing reveals the cluster gravitating mass regardless of the matter
compositions and their dynamical status, the consistency between these
three mass estimators indicates that clusters may be regarded as
dynamically-relaxed systems as a whole. Further support to this argument
comes from the study of the correlations between the optically determined
velocity dispersion of cluster galaxies, the X-ray measured luminosity and
the temperature. Namely, although many X-ray clusters show irregular
configurations and are substructured, there are tight correlations between
the velocity dispersion of optical galaxies,  the temperature and
the luminosity of the intracluster gas (Evrard, Metzler, \& Navarro 1996;
Mohr, \& Evrard 1997; Wu, Fang \& Xu 1998; Wu, Xue \& Fang 1999; Mohr, 
Mathiesen, \& Evrard 1999).

In a word, the observational evidences for whether galaxy clusters are 
virialized seem to be contradictory. In this paper, we intend to highlight
this contradiction. We shall show that beyond the scale of full 
virialization there exists a radius range in which clusters are 
quasi-virialized, i.e. the internal structure of {\it individual} cluster 
exhibits a substantial departure from dynamical relaxation, while many
{\it statistical} properties of clusters are approximately the same as
those of the virialized system. For instance, the virial relation 
$\sigma_v \propto M^{1/2}$, where $\sigma_v$ is the velocity dispersion and 
$M$ is the mass of a system within a given radius, is a good approximation on
scales larger than virial radius, even though many clusters are irregular in
shape and show substructures on these scales. Actually, such a ``duality'' 
is an
indicator of the transition from pre-virialization to virialization. This
transition can roughly be described by three stages: a) full virialization,
in which both the 
individual objects and their statistical properties are virialized;
b) quasi-virialization, in which the ensemble statistical features are
similar to those of the virialized systems, but the scattering from the
averages is still significant; c) pre-virialization, in which both individual
objects and their ensemble statistical features are significantly different
from virialization.

In hierarchical clustering scenario, the larger the scale, the less the
virialization. Thus, the dynamical evolution from phase a) to b) or from
phase b) to c) can be revealed by studying the dynamical properties of
the objects on the scale of clusters and beyond. In particular, one can
address the question: On what scale will the transition of phase a)
to b) or phase b) to c) occur? 
We are interested in
this issue also because the virialization evolution is model-dependent.
It is hoped that the scale-dependence of virialization is useful
for discrimination of various cosmological models.

The present paper is organized as follows. In \S 2, we will describe 
our samples of cluster halos which are multi-scale identified from N-body 
simulations under three different cosmological models: 1) the standard cold
dark matter (SCDM), 2) low density, flat CDM model with a non-zero
$\lambda$ (LCDM), and 3) the open CDM (OCDM) model. The Press-Schechter (PS) 
formalism is also introduced so that the simulation results can be confirmed
by a semi-analytical approach. We then investigate in \S 3 various
statistical indicators of virialization, including density
contrast, configuration, substructure and velocity dispersion. The emphasis
is to find the scale-dependence of these features, and determine the scale
range of virialization and quasi-virialization. In \S 4 we will
discuss the applicable range of the virial mass estimator as a statistical 
measure for quasi-virialized systems. In \S 5 the temperature function of 
clusters will be constructed and used for model discrimination.  Finally, 
we will briefly summarize our conclusions in \S 6.

\section{Simulated samples of $r_{cl}$-clusters}

\subsection{Samples}

The samples to be used for analyzing the scale-dependence of virialization
are produced by N-body simulations with the P$^3$M code developed by
Y.P.Jing (Jing \& Fang 1994; Jing et al. 1995). We employed the following
cosmological models: 1) SCDM, 2) LCDM, and 3) OCDM. The cosmological 
parameters ($\Omega_M, \Omega_{\Lambda}, h, \sigma_8$) are taken to be 
(1.0,0.0,0.5,0.62), (0.3,0.7,0.75,1.0), and (0.3,0.0,0.75,1.0) for the SCDM,
LCDM and OCDM, respectively. These parameters provide a consistent
description for many observational properties of the universe, especially 
the abundance of clusters (e.g. Jing \& Fang 1994; Bahcall, Fan \& Cen 
1997). 

Other parameters in our simulations, i.e. box size $L$, number of
particles $N_p$ and the effective force resolution $\eta$, are chosen to
be ($L$,$N_p$,$\eta)= (310 h^{-1}$ Mpc, $64^3,0.24 h^{-1}$ Mpc). We have 
run 8 realizations for each model. A particle has mass of
$3.14\times 10^{13}\Omega_M \  h^{-1}$ M$_{\odot}$, which is small enough to
resolve reliably the rich clusters of $M > 3.0 \times 10^{14} h^{-1}$
M$_{\odot}$. We have also performed simulations with a COBE normalization
$\sigma_8=1$ for SCDM. Three realizations were made.

\subsection{Identification of $r_{cl}$-clusters}

There are two ways of studying the scale-dependence of dynamical status
of clusters. One is to identify clusters with a given size, like Abell
radius $r_{ab}= 1.5$ $h^{-1}$ Mpc, and then study the dependence of
dynamical properties of these clusters on scales less or greater than 
$r_{ab}$. Another is to identify clusters by multi-scale resolution analysis,
and produce samples of $r_{cl}$-clusters, i.e. clusters identified with
scales $r_{cl}$, and then, study the $r_{cl}$-dependencies, and also the
radius-dependence of $r_{cl}$-clusters.

If all objects on all scales $r_{cl}$ are virialized, $r_{cl}$-clusters
should show ``clouds in clouds" morphology, i.e. all smaller $r_{cl}$
clusters locate in the centers of larger $r_{cl}$ clusters. In this case, 
the multi-scale identification will actually 
produce the same samples as the
Abell radius identification. The $r_{ab}$-identification and the multi-scale
identification would be essentially equivalent. However, in the evolutionary
stage of pre-virialization, the two identifications are different from
each other. The multi-scale identification is more effective to study
problems of scale-dependence of clusters and beyond (Xu, Fang \& Wu 1998;
Xu, Fang \& Deng 1999; Fang \& Xu 1999).

The detailed procedure of the multi-scale identifications based on the
discrete wavelet transform (DWT) has been presented in the papers of Xu,
Fang \& Wu (1998) and Xu, Fang \& Deng (1999). The technique of applying
the DWT for large scale structure study is reviewed in Fang \& Thews
(1998). Briefly, we first describe the distribution of the particles by a
3-D matrix, and then do the fast 3-D {\it Daubechies 4} DWT and the
reversed transformations to find the wavelet function coefficients (WFCs)
and scaling function coefficients (SFCs) on various scales. For each scale,
the cells with SFCs larger than those of the random sample by a given
statistical significance, say 3$\sigma$, are picked up as the halos of
cluster candidates. Around each of the candidates, we place a $6^3$ grid
with the size of cluster diameter and search for the accurate center. The
cluster center is taken as the position where the largest mass is surrounded.
The mass $M$ of a cluster is measured by counting the particles within a
sphere of radius $r_{cl}$. It is enough to take $r_{cl}$ to be numbers
of bi-fold, i.e. 0.75, 1.5, 3, 6, 12 and 24 $h^{-1}$Mpc. Since the $r_{cl}$
sphere does not depend on the shape of the wavelet functions, this
identification is independent of the selection of wavelet function.

With the identified $r_{cl}$-clusters, one can find the mass function (MF) or
abundance $n(>M,r_{cl})$, which is the number density of $r_{cl}$-clusters 
with mass larger than $M$. As an example the MFs of  $r_{cl}$-clusters of the
SCDM model are ploted in Fig. 1. It shows scaling behavior. For other models 
the MFs show  similar behaviors (Fang \& Xu 1999). The abundance 
$n(>M,r_{ab})$ of clusters on the Abell radius given by the DWT
identification is found to be the same as those given by usual
friends-of-friends (FOF) identification (Xu, Fang \& Wu 1998;
Xu, Fang \& Deng 1999).

Since each halo or cluster is characterized by two
parameters $M$ and $r_{cl}$, it is inconvenient to define richness by its
mass alone. Instead, the richness can be defined by the abundance,
$n(>M,r_{cl})$, or by the mean separation of neighbor clusters
%eq1
\begin{equation}
d=\frac{1}{n(>M,r_{cl})^{1/3}}.
\end{equation}
The richness of $r_1$-clusters with masses $>M_1$ is equal to the richness
of $r_2$- clusters with masses $>M_2$ if their $d$'s are the same.
For instance, for the top 25 clusters on various scale $r_{cl}$ in the
simulation box 310$^3$ $h^{-3}$ Mpc$^{3}$, their richness is
$d=310/25^{1/3}\simeq 106$ $h^{-1}$ Mpc. For a given $r_{cl}$, this
richness definition has its usual meaning, i.e. the larger the $M$,
the less the $n(>M,r_{cl})$ and the higher the richness.

\subsection{The Press-Schechter formalism for $r_{cl}$-clusters}

The MFs given by the N-body simulation can be explained by the 
hierarchical clustering of cosmic structures. In the models of hierarchical 
clustering with Gaussian initial perturbations, the comoving number density 
of collapsed halos at $z=0$ in the mass range $M$ to $M+dM$ is given by
(Press \& Schechter 1974)
%eq2
\begin{equation}
N(M) dM= -\sqrt{\frac{2}{\pi}}\frac{\rho_0}{M^2}
        \frac{\delta_c}{\sigma(M)} \exp
        \left( \frac{-\delta_c^2}{2\sigma(M)^2} \right)
\frac{d\ln \sigma(M)}{d \ln M} dM,
\end{equation}
where $\delta_c\approx1.69$ almost independent of cosmologies.
$\sigma(M)$ is the linear theory $rms$ mass density fluctuation
in spheres of mass $M$ at redshift $z=0$ within a top-hat window of radius
$R$, and is determined by the initial density spectrum $P(k)$ and
normalization factor $\sigma_{8}=\Delta(8h^{-1}$Mpc,0).
 
 From eq.(2), the cumulative number density of the collapsed halos with 
mass greater than $M$ is given by
%eq3
\begin{equation}
n(>M)=\int^{\infty}_{M} N(M)dM.
\end{equation}
Equations (2),(3) have been found to be good descriptions of the mass 
functions of collapsed clusters identified by FOF method from N-body 
simulation (e.g. Mo, Jing \& White 1997). Figure 1 shows that
PS predications fit very well with the 
mass function (MF) of clusters on the Abell radius given by the DWT 
identification.

As pointed out by Mo \& White (1995) and Bi \& Fang (1996; 1997), 
the PS formalism can be employed to approximately describe not only  
collapsed halos, but also uncollapsed regions, i.e. both collapsed halos 
and uncollapsed regions can be viewed as the sum of various individual
top-hat spheres. An uncollapsed region (or ``quasi-virialized halo" in 
terminology of this paper) corresponds to a region in which the initial 
linear fluctuation is less than the threshold 1.686 at the redshift 
considered. As a consequence, one can consider the initial density field 
as a system consisting of many spheres, each of which has initial radius 
$R$ and density contrasts $\delta_0$, and mass  is equal to
$M \equiv \frac{4}{3}\pi R^3 \rho_0$.

Let's consider an arbitrary spherical volume of radius $r_{cl}$ 
at redshift $z=0$. Matter in this volume can come from various initial 
spheres with $R$ and $\delta_0$. For a given $r_{cl}$, one can find the 
relationship between $R$ and $\delta_0$ by considering the nonlinear 
evolution of a massive shell under spherical symmetry (Padmanabhan 1993;
Mo \& White 1995; Bi \& Fang 1996). 
Then, we have the relation
%eq4
\begin{equation}
\frac{\delta_{th}}{\delta_c} = f(\frac{r_{cl}}{R}).
\end{equation}
It means that a spherical region of $\delta_{th}$ and $R$ will evolve 
into $r_{cl}$  at $z=0$, which includes both collapsed and uncollapsed 
objects. The function $f(x)$ is given in Bi \& Fang (1996). 

If we identify clusters with radius $r_{cl}$ in an arbitrary 
spatial domain $\Delta x $, the fraction of the total mass 
$\rho_0 \Delta x$ contained in clusters of mass $>M$ corresponds to the
fraction of mass which have initial fluctuations
larger than a given $\delta _{th}$. For a Gaussian field,
this fraction is given by
%eq5
\begin{equation}
F(>M,r_{cl})=\int ^\infty _{\delta _{th}(M,r_{cl})}
 \frac{1}{\sqrt{2\pi} \sigma _{M}}
      \exp{\left(-\frac{\delta ^2}{2\sigma _{M} ^2}\right)} d\delta,
\end{equation}
where $\sigma_{M}^2$ is the variance of density perturbation on
scale $R$. Because a spherical region of $\delta_{th}$ and $R$ will
evolve into $r_{cl}$ at $z=0$,
initial spherical regions of $\delta \ge \delta _{th}$
and $R$ will evolve into radii less than $r_{cl}$ at $z=0$. Therefore, the
comoving number density of the collapsed and uncollapsed PS spheres with 
mass in $M \to M + dM$, which are spheres identified with radius $r_{cl}$ 
at $z=0$, is given by
%eq6
\begin{eqnarray}
%\begin{array}{ll}
N(M,r_{cl}) dM & = &  -2\frac{\rho_0}{M} \frac{\partial}{\partial M}
   {\rm erfc}\left[\frac{\delta_{th}}{\sqrt{2}\sigma (R)}\right] dM
\\ \nonumber
 &  = & -\sqrt{\frac{2}{\pi}} \frac{\rho_0}{M^2}
\nu e^{-\frac{\nu^2}{2}}(\frac{d ln\sigma}{d lnM}
-\frac{d ln\delta}{d lnM}),
%\end{array}
\end{eqnarray}
where $\nu \equiv \delta(R,r_{cl})/\sigma(R)$
denotes the relative height of the initial perturbation (evaluated 
at redshift $z=0$). The number 2 in the equation is the notorious
$ad \ hoc$ multiplication factor of PS formalism
(see discussion in Bond et al. 1991).

The cumulative number density, $n(>M,r_{cl})$, of collapsed and 
uncollapsed $r_{cl}$-halos with mass greater than $M$ should be
%eq7
\begin{eqnarray}
%\begin{array}{ll}
n(>M,r_{cl}) & = & \int ^{\infty} _M N(M,r_{cl}) dM \\ \nonumber
 &	= &-\frac{3}{(2\pi)^{\frac32}}\int ^{\infty} _R
	\nu e^{-\frac{\nu^2}{2}} \frac{1}{x^3} (
	\frac{d ln\sigma}{dx}-\frac{d ln f}{dx})dx.
%\end{array}
\end{eqnarray}
These MFs are also plotted in Fig. 1. It shows that the PS formalism  
basically explains the mass functions given by N-body simulation,  
especially the scaling behaviors. Meanwhile, Fig. 1 shows that the 
PS calculations systematically underpredict the MFs on scale 
$r_{cl} > 3$ h$^{-1}$ Mpc. This is most likely caused by the hypothesis 
of spherical symmetry of the dynamic evolution of uncollapsed regions. 
A similar discrepancy has also been found in the two-point correlation 
of halos in underdense (or uncollapsed) regions, i.e. the PS formalism 
significantly underpredicts the halo-halo correlation when small halos 
are in low-density environments (Jing 1998). Nevertheless, the PS result 
shows the scale-dependence of statistical features given by the PS
formalism is the same as that of N-body simulation. 
This agreement is very useful to reveal 
the status of quasi-virialized (or uncollapsed) regions.

\section{Virialization and quasi-virialization}

\subsection{Density contrasts of $r_{cl}$-clusters}

A popular indicator of virialization of clusters is the density contrast.
For spherical collapse in a standard flat universe, the virial radius is
usually measured as the radius within which the mean density contrast is
$\sim 178$, i.e. $\rho_{vir}/\rho_b=178$. For open universe, the density
contrast within virial radius is in the range of $\sim 100/\Omega_M$
to $200/\Omega_M$ (Lacey \& Cole 1993). For models of SCDM, LCDM and OCDM,
the density contrasts $\rho_{vir}/\rho_b$ are equal to 178, 335, 402,
respectively.

Observationally, the spatial range of virialization is characterized by
$r_{200}$, the radius where the average interior number density of galaxies
is $200$ times higher than the average. Thus, the deviation from
virialization of a $r_{cl}$-cluster can be quantified by the mean mass
contrast, $\langle \Delta M \rangle /\langle M \rangle$, where
$\langle M \rangle$ is the mean mass within radius $r_{cl}$, and
$\langle \Delta M \rangle$ the mean mass excess of the $r_{cl}$-clusters.
Fig. 2 plots the mass contrasts of $r_{cl}$-clusters with richness
$d=$ 30, 50 and 100 $h^{-1}$ Mpc for the SCDM, LCDM and OCDM models,
respectively. The lines in LCDM almost overlap with those for OCDM. As 
expected, the more massive (i.e. the larger $d$) the clusters, the closer
to virialization. For the SCDM clusters with richness $d=$ 30, 50, 
and 100 $h^{-1}$ Mpc, the virialized radii $r_{vir}$ at $z=0$ are 1.09, 
1.33, and 1.60 $h^{-1}$ Mpc, respectively. They are 1.03, 1.33, 1.82
$h^{-1}$ Mpc for LCDM, and 0.96, 1.27, 1.70 $h^{-1}$ Mpc for OCDM.

Therefore, except for the richest clusters, the full virialization generally
is realized only on scales less than $r_{ab}= 1.5$ $h^{-1}$ Mpc. This is
consistent with the observed result that $r_{200}$ is usually smaller than
the Abell radius $r_{ab}$. Namely, except for the core regions of clusters,
Abell clusters as a whole are not yet fully virialized.

In the case of positive bias, $r_{200}$ given by galaxies is larger
than that of dark matter, and therefore, the real virialized range will
be even smaller than that given by galaxy observations. As a result, in the
scale range of equal to and larger than the Abell radius, halos for
$r_{cl}$-clusters are mostly in the stages of quasi- or even 
pre-virialization.

\subsection{Configurations of $r_{cl}$-clusters}

The spatial configuration of clusters is also a useful indicator of 
virialization. As discussed above, if the halos of clusters are 
fully virialized, their spatial configurations have to be regular ``clouds 
in clouds", i.e. the smaller $r_{cl}$-clusters are located in the centers 
of larger $r_{cl}$-clusters.

Fig. 3 plots a projected distribution of $r_{cl}$-clusters identified
from one realization of model OCDM within box 310 $h^{-3}$ Mpc$^3$, which 
illustrates the top 25 massive clusters for each $r_{cl}$ (the richness 
is $d= 106$ $h^{-1}$ Mpc). The scales $r_{cl}$ are taken to be 
$0.75,\ 1.5,\ 3, \ 6, \ 12$ and 24  $h^{-1}$Mpc, respectively.
Fig. 3 shows a variety of configurations of the halos. Apparently, 
some halos show ``clouds in clouds" morphology, while some do not.

According to the scale of $r_{200}$ in Fig. 2, it is expected
that, for such a high richness, clusters of $r_{cl} \leq $ 2 $h^{-1}$ Mpc 
have to be symmetrical and concentric ``clouds in clouds" structures. 
Fig. 3, indeed, shows that 21 of the 25 $r_{cl}=0.75$ and 1.5 $h^{-1}$ Mpc 
are concentric. These $r_{cl}=$ 0.75-1.5 ``clouds in clouds" objects 
are fully virialized.

On the other hand, there are seven $r_{cl}$ = 3 $h^{-1}$ Mpc clusters which
do not contain either $r_{cl}=0.75$ $h^{-1}$ Mpc or 1.5 $h^{-1}$ Mpc 
clusters in their centers. These are massive clusters on scale
3 $h^{-1}$ Mpc, but without fully virialized cores. On scales
$r_{cl}\geq 3$ $h^{-1}$ Mpc, more $r_{cl}$-clusters are asymmetric,
irregular, and without small-scale high density peaks within them. These
features show that identifying structures with a given scale like $r_{ab}$
may miss massive objects which do not contain high density peaks on scale
$r_{ab}$. Only in the range $r_{cl} \leq r_{200}$, the multi-scale
identifications find the same sample as the single-scale identification.

In terms of mass and length scale, $r_{cl}=24$ $h^{-1}$ Mpc clusters actually 
are superclusters. Observationally, supercluster is defined as cluster of 
rich clusters (top $r_{ab}$-clusters). In Fig. 3, one can see that some 
$r_{ab}$-clusters are located together to form larger scale clumps or 
filaments. They are obviously superclusters. Fig.3 shows, however, that some 
of the top 25 $r_{cl}=24$ $h^{-1}$ Mpc halos contain nothing of 
$r_{cl} \leq $ 24 $h^{-1}$ Mpc massive halos. This is, although the mass
and length scale of these objects are the same as superclusters, they do not
contain any of the top 25 clusters on smaller scales. This result indicates
that the identification by ``cluster of rich clusters'' probably is
incomplete. Recently, it has been found that the dynamical properties of 
superclusters are not only determined by rich clusters, but also by numerous 
galaxies in the intra-cluster regions (Small et al. 1998). Therefore,
to study the dynamics on scales beyond clusters, one should directly identify
structures from matter or galaxy distributions, instead of using only rich
clusters.

\subsection{Substructures and potential minimums of $r_{cl}$-clusters}

Similar to the configuration, the number of substructures in a cluster is 
also an effective measure of the deviation from virialization. The asymmetry 
of the configuration of clusters actually is caused by their substructures. 

Considering that the time scale required for a sound wave in the intracluster
gas to cross a cluster is shorter than the dynamical time of the cluster, it 
is reasonable to assume that the intracluster gas is in hydrostatic 
equilibrium with local gravitational potential, and therefore, X-ray images
map the cluster potential locally (Jing et al. 1995). Thus, one can identify
substructure as potential minimum of clusters.

We search for the potential minimums by the same algorithm of Jing et al.
(1995). Briefly, we first place a mesh with 0.24 $h^{-1}$Mpc resolution
around each center of identified $r_{cl}$-cluster halos. The size of this
mesh is taken to be 7.5, 7.5, 7.5, 15 and 30 $h^{-1}$Mpc for clusters of
$r_{cl}=$ 0.75, 1.5, 3, 6 and 12 $h^{-1}$Mpc, corresponding to grid points
$32^3,\ 32^3,\ 32^3,\ 64^3$ and $128^3$, respectively. We accumulate mass
density values for the meshes from particles. The mass of each particle is
smoothed by a Gaussian kernel
%eq8
\begin{equation}
W(r,s)={1\over (2\pi)^{3/2} s^3} \exp{-r^2\over 2s^2},
\end{equation}
where $s$ is the smoothing length. The smoothing length $s_i$ of particle 
$i$ is equal to the local mean separation $d_i$, calculated by counting five 
nearest particles around $i$. Then the density value on arbitrary cell $j$ 
is given by
%eq9
\begin{equation}
\rho_j=\sum_i m_i W(r_{ji},s_i),
\end{equation}
where $r_{ij}$ is the separation between cell $j$ and particle $i$ of 
mass $m_i$. The summation in Eq.(9) is made over all particles within
radius 3.75 h$^{-1}$ Mpc. With this mass distribution, gravitational
potential on the grid can be obtained. We then identify a cell to be
a potential minimum if its potential value is smaller than those of its all
26 neighbors.

Obviously, if a $r_{cl}$-cluster contains only one potential minimum, it
means no substructure. Fig. 4 shows the dependence of average number of
substructures per $r_{cl}$-cluster on the count of the particles within the
cluster, which is proportional to the cluster mass. As expected, the larger
$r_{cl}$ clusters contain more substructures
on average because larger scale objects are less virialized. 

A ``surprising''  result shown in Fig. 4 is that for  
$r_{cl} > 3 \ h^{-1}$ Mpc clusters, the more massive the clusters (i.e. the
more the count of particles), the more the substructures. This seems
to be contradicting with the general conclusion of last section: the more
massive the cluster, the closer to virialization. Actually, this is due to
the threshold used for identifying substructures. In massive objects
substructures with deep potential valleys are easy to grow up, but not so
for less massive objects. For clusters with
$r_{cl} = 1.5\ h^{-1}$Mpc, containing no substructures on average (as shown
in Fig. 4) doesn't mean that all $r_{cl} =$ 1.5 $h^{-1}$Mpc clusters are
fully virialized. Contrarily, low richness $r_{cl} =$ 1.5 $h^{-1}$Mpc
clusters are pre-virialized (Fig. 2). They simply do not have enough mass
to form substructures above the defined threshold.

Similar phenomena can be seen from Fig. 5 which shows the probability 
distributions of the number of substructures per $r_{cl} = 24\ h^{-1}$Mpc
clusters with richness $d= 80\ h^{-1}$ Mpc, i.e. the top 59 richest clusters
in the simulation box 310$^3$ $h^{-3}$ Mpc$^{3}$. From Fig. 5 it is seen 
that the SCDM clusters generally are less substructured than those of the
LCDM and OCDM. It doesn't mean that the SCDM clusters are more close to
virialization. Instead, the deep potential valleys for substructures in the
SCDM $r_{cl} = 24\ h^{-1}$Mpc clusters formed later than the LCDM and OCDM.

Fig. 6 gives the $r_{cl}$-dependence of the mean number of substructures
per $r_{cl}$-clusters having richness $d=$ 80\ h$^{-1}$ Mpc.  For all
the three cosmological models, the number of substructures per
$r_{cl}$-clusters increases with $r_{cl}$. As Fig. 5, the mean number of
substructures in model SCDM is less than those of LCDM and OCDM. The
substructure number at redshift $z=0.5$ is less than that of $z=0$, too.
This is once again to indicate that deep potential valleys have not yet
developed at $z=0.5$.

Thus, despite 
substructure is an indicator of the deviation from 
virialization, there is no simple relation between virialization and 
substructures. A system with rich substructures definitely deviates from 
virialization, but the existence of substructures is not a necessary 
condition of the deviation from virialization. Systems with more deep 
potential valleys are not always in a lower degree of virialization than 
those with fewer potential valleys. This point is important in studying 
objects on scales larger than $r_{ab}$.
 
\subsection{Velocity dispersion}

The most decisive measure of virialization certainly is the velocity 
dispersion of a cluster. For a virialized $r_{cl}$-cluster we have
$\sigma_v^2 \propto M(<r_{cl})$, in which $M(<r_{cl})$ denotes the mass of
a cluster within radius $r_{cl}$. The velocity dispersion $\sigma_v$ of
$r_{cl}$-cluster is calculated from the velocities of all CDM particles
within radius $r_{cl}$ with respect to the center of the cluster. Fig. 7
shows the three dimensional $\sigma_v$ of $r_{cl}= 0.75 - 24\ h^{-1}$ Mpc
clusters identified from three realizations of the OCDM model at $z=0$.

In the case of full virialization, $\sigma_v$ of the system with a given
size $r_{cl}$ should be completely determined by its masses $M(<r_{cl})$.
Fig. 7 shows that $\sigma_v$ is almost completely determined by $r_{cl}$
and $M(<r_{cl})$ if clusters are as massive as
$M(<r_{cl}) >$ $10^{14.7},\ 10^{15}$ and $10^{15.2} h^{-1}M_{\odot}$, 
respectively for $r_{cl} =$ 0.75, 1.5 and 3 $h^{-1}$ Mpc. Namely, in these 
parameter ranges ($r_{cl}$, $M$), clusters are fully virialized. Out of it,
velocity dispersion $\sigma_v$ is no longer completely determined by radius
$r_{cl}$ and mass $M(<r_{cl})$. As shown in Fig. 7, the scattering of 
velocity dispersion $\sigma_v$ around their mean $\bar{\sigma}_v$ can be 
as large as 15\% (see the vertical error bars of Fig. 7). This scattering 
is not caused by Poisson fluctuations. For $r_{cl} = 0.75 - 3 h^{-1}$ Mpc, 
the scattering increases with the increase of $r_{cl}$ and the decrease of 
$M(<r_{cl})$. Therefore, the scattering is due to the deviation from 
virialization.

The mean velocity dispersion $\bar{\sigma}_v$ of the
panels of $r_{cl} \leq 6$ $h^{-1}$ Mpc in Fig. 7 is found to be proportional
to $M^{1/2}(<r_{cl})$. Fig. 8 plots the result of
$\gamma=d \overline{\log \sigma}_v/d \log M(<r_{cl})$ for the data
set shown in Fig. 7.  Fig. 8 clearly demonstrates that
for scales $r_{cl} \leq 6$ $h^{-1}$ Mpc and richness $d =$ 20 -- 100 $h^{-1}$
Mpc, all the three models yield $\gamma \simeq 0.5$, which is the value for
virialized systems. Namely, the ensembly averaged relations between
mass and velocity dispersion of $r_{cl} \leq 6$ $h^{-1}$ Mpc clusters are
the same as those for the virialized systems. Even for ensemble of rich
$r_{cl}=12$ $h^{-1}$ Mpc clusters with $d \geq 80$ $h^{-1}$ Mpc,
virialization holds approximately true. However, this approximation
is no longer valid for the $r_{cl}=24$ $h^{-1}$
and $d \leq 100$ $h^{-1}$ Mpc clusters.

Strictly speaking, virial theorem is $K=W/2$, where $W$ and $K$ represent,
respectively, the potential and kinetic energy of the system. 
Therefore, it includes two numbers 1.) the index $\gamma = 0.5$ from 
$K \propto W$; 2. the coefficient $1/2$ from $K/W =1/2$. 
In the above used parameter space of $r_{cl}$ and $d$, both the index and the
coefficient are found to be not different from 1/2 by 10\%. One can define 
``quasi-virialization" to be a dynamically evolutionary stage of
$r_{cl}$-clusters, for which {\it ensemble averaged} properties 
approximately agree with those required by virialization, while
{\it individual} clusters may significantly depart from virialization. 

Thus, Figs. 7 and 8 show that the quasi-virialization range for
the LCDM and OCDM clusters at $z=0$ is from radius $r_{vir}$ to about
6 $h^{-1}$ Mpc with richness $d > 20$ $h^{-1}$ Mpc, and to about 
12 $h^{-1}$ Mpc with $d \geq 80$ $h^{-1}$ Mpc. For the SCDM clusters of 
$r_{cl}$ $\sim$ 12 $h^{-1}$ Mpc, only the richest clusters 
($d \geq 100$ $h^{-1}$ Mpc) are quasi-virialized. The value $\gamma$ 
at redshift $z=0.5$ is also plotted in Fig. 8. For the LCDM and OCDM, the 
quasi-virialization is already realized for structures of 
$r_{cl}< 6 h^{-1}$Mpc at $z=0.5$. For the SCDM, the  quasi-virialization 
approximately holds for clusters of $r_{cl}$ $\leq$ 6 $h^{-1}$ Mpc, 
while $r_{cl}=12$ $h^{-1}$ Mpc clusters are completely out of the
quasi-virialization range.

The redshift evolution of $\gamma$ for models SCDM, LCDM and OCDM is
shown in Fig. 9, which gives the mean $\gamma$ at redshifts
$z=$ 0, 0.2, 0.5, 0.8 for samples of top 59 clusters, or richness
$d=$ 80 $h^{-1}$ Mpc, where $r_{cl}$ is taken to be 1.5, 12 and 24
$h^{-1}$ Mpc. For clusters of $r_{cl} \leq$ 12 $h^{-1}$ Mpc in the LCDM and
OCDM models, we have $\gamma \simeq 0.5$ in the redshift range of
$z\leq 0.8$. But in the SCDM model, the stage of quasi-virialization
has not yet been reached at redshift $z > 0$ for clusters of 
$r_{cl}=$ 12 $h^{-1}$ Mpc.
 
\section{The virial mass estimator}

With the above analysis on quasi-virialization, one can now study the
applicability of the virial mass estimator. Especially, the 
individual-statistical (or individual- ensemble) duality of 
quasi-virialization is very useful for determining the available range 
of the virial mass as a statistical measure.

The virial estimator of the total mass of a cluster within radius $r$ 
is given by (e.g. Bahcall \& Sarazin 1977; Mathews 1978),
%eq10
\begin{equation}
M_v(<r)=-\frac{\sigma^2_v r}{3G}\left(\frac{d\ln \rho_{gal}}{d\ln r}
 + \frac{d \ln \sigma^2_v}{d \ln r}\right )\simeq
\frac{\sigma^2 }{3G}f(r),
\end{equation}
where $\rho_{gal}$ is the density of galaxies. For isothermal sphere 
approximation of the galaxy distribution, we have 
$f(r)=3r^3/(r^2+r_c^2)$, 
where  $r_c$ is the core radius. Basically, the applicability of eq.(10)
relys on the following three assumptions: 1.) The second term in the
bracket of eq.(10), i.e. $d \ln \sigma^2_v/d \ln r$, is negligible. 
Namely, $\sigma^2_v$ should be independent of $r$; 2.) For a given $r$, the
velocity dispersion 
$\sigma^2_v$ is proportional to $M(<r)$,
i.e. $\gamma_v\equiv d\ln \sigma_v/d\ln M=0.5$; 3.) The profile of mass
distribution $f(r)$ doesn't significantly depend on $\sigma_v$ and richness 
$d$.

The assumption 2 has been studied in \S 3.4. The mean $\gamma$ is indeed
equal to 0.5 for quasi-virialized clusters. We will show below that
the assumptions 1 and 3 are also statistically available for quasi-virialized 
systems.

\subsection{Profiles of velocity dispersion}

To test the assumption 1, we calculated the mean velocity dispersion 
$\sigma_v$ within radius
$r$ among the 50 most massive clusters (or richness $d \sim $ 85 $h^{-1}$ Mpc)
at scales $r_{cl}=$ 0.75, 1.5, 3, 6, 12 and 24 $h^{-1}$ Mpc. The results
averaged from 8 realizations for OCDM and SCDM are plotted in Fig. 10.
The LCDM gives similar results as the OCDM. Fig. 10 shows that the profiles of
velocity dispersion of $r_{cl} \leq 6 $  $h^{-1}$ Mpc clusters are rather
flat, and the variations of the velocity profiles over the radius range
$0.5 < r \leq 10$ $h^{-1}$ Mpc are no more than $20\%$. Beyond $
r=10$ $h^{-1}$ Mpc, the velocity of particles is dominated by the Hubble
expansion.

Fig. 10 also shows that $\sigma_v$ is virtually $r_{cl}$-independent for
clusters of $r_{cl}\leq 3$ $h^{-1}$ Mpc.  When $r_{cl} > 3$ $h^{-1}$ Mpc,
$\sigma_v$ is lower for larger $r_{cl}$. However, the velocity dispersion 
of $r_{cl} = 6$ $h^{-1}$ Mpc clusters is only $\sim$ 20\% lower than that
of $r_{cl} \leq 3$ $h^{-1}$ Mpc clusters. Even in the case of 
$r_{cl}= 12$ $h^{-1}$ Mpc, the $\sigma_v$ is not different from
$r_{cl} \leq 3$ $h^{-1}$ Mpc clusters by a factor of 2.

Despite many top $r_{cl}=$ 6 and 12 $h^{-1}$ Mpc clusters do not contain
top $r_{cl}< 6h^{-1}$ Mpc clusters as ``clouds in clouds", the {\it mean} 
profile of the velocity dispersion $\sigma_v$ remains to be flat from
1 to $\sim$ 10 $h^{-1}$ Mpc for clusters of $r_{cl} \leq $ 6 $h^{-1}$ Mpc
and $d >$ 40 $h^{-1}$ Mpc or $r_{cl} \leq $ 12 $h^{-1}$ Mpc and 
$d >$ 80 $h^{-1}$ Mpc. Therefore, the assumption 1 is statistically correct
for quasi-virialized clusters.

\subsection{Profile of mass distributions}

To test the assumption 3, we calculated the cross-correlation function 
between clusters and mass particles, $\xi_{cm}(r) = N(r)/N_{exp}(r)-1$, 
where $N(r)$ is the number of particles within a spherical shell with radius 
$\sim r$, and $N_{exp}(r)$ is the mean number expected from random 
distribution. 

Fig. 11 gives the cluster-particle cross-correlation functions of 
clusters with $r_{cl}=$ 1.5, 6 and 24 $h^{-1}$ Mpc and richness 
$d=90$ $h^{-1}$ Mpc. It shows that the mean cross correlation functions 
$\xi_{cm}(r)$ are about the same for all $r_{cl}$-clusters.
 One can describe the correlation function $\xi_{cm}$ as 
$\xi_{cm} \propto r^{\alpha-3}$ (or,
$f(r) \propto r^{\alpha}$) in the radius range from 1 to 20 $h^{-1}$ Mpc. 
The index $\alpha$ and their variance of each $r_{cl}$ are listed in 
Table 1. For the LCDM and OCDM, all the indexes of  
$r_{cl} \leq$ 6 $h^{-1}$ Mpc clusters are $\alpha \simeq 0.5$. 
For the SCDM $\alpha$ is slightly lower than 0.5 at 
$r_{cl} \leq$ 6 $h^{-1}$ Mpc.

Fig. 12 plots $\xi_{cm}(r)$ for the $r_{cl}=$ 6 $h^{-1}$Mpc
clusters with richness from $d= 40$ to 120 $h^{-1}$ Mpc in the OCDM. 
$\xi_{cm}(r)$ shows a systematic increase with $d$, indicating that the
mean mass of clusters increases with $d$. However, the shapes of density
profiles for all richness can approximately be written as
$\xi_{cm}(r) \propto r^{\alpha-3}$, or $f(r)\propto r^{\alpha}$, and 
$\alpha \simeq 0.5$ in the radius range 1 - 20 $h^{-1}$Mpc. That is, the 
shape of mass profile is independent of the richness. 

Moreover, for a given $r_{cl}$ such as $r_{cl}$= 6 $h^{-1}$ Mpc,
the mean velocity dispersion $\sigma$ increases with mass or $d$ (Fig. 7). 
Therefore, the cluster-particle cross-correlation functions $\xi_{cm}(r)$ 
with different $d$ in Fig. 12 actually is for different $\sigma$.
Therefore, Fig. 12 also shows that the shape of mass profile also does not 
depend on the mean velocity dispersion.

Thus, one can conclude that the mass profile
$f(r)\propto r^{\alpha}$ with $\alpha \simeq 0.5$ is weakly dependent
on $r_{cl}$, richness and velocity dispersion. One may safely employ the 
assumption 3 at least on scales $\leq 6$ $h^{-1}$ Mpc, i.e. in the range of 
quasi-virialization. Actually, the independence of the cluster mass profiles 
on the richness and velocity dispersion has been found by Jing et al. (1995).

It should be pointed out that the radius range shown in Figs. 11 and  12
are generally larger or much larger than $r_{200}$. That is, we study only
on scales much larger than the cores of clusters, and therefore, the
core radius is less important.

\subsection{The goodness of the virial mass estimator}

Accordingly, mass estimator eq.(10) is good as a statistical measurement
for clusters on radius larger than $r_{vir}$ or $r_{200}$, and within the
range of quasi-virialization. The
goodness of this statistical measurement can be seen from Fig. 7, in which 
the vertical error bars show the deviation of $\log \sigma_v$ from their 
virialization average. The error bars $\Delta \log \sigma_v$ are of
the order of $\sim 0.15$. Thus, $M_v$ given by estimator (10) should have
an uncertainty of about $\Delta M_v/M_v \sim 0.30$. This result provides
an explanation for the fact that the cluster virial masses determined from
optical galaxies statistically are in a good agreement with gravitational
lensing-derived masses on scales much greater than the core radii. For
instance, based on a sample of lensing and optical clusters, the
correlations between the virial mass $M_{v}$ and weak lensing-derived
mass $M_{len}$ of clusters in the scale range of $0.25 -  1.5$ $h^{-1}$ Mpc
are found to be (Wu \& Fang 1997; Wu et al. 1998)
%eq11
\begin{equation}
M_{len}= (0.63 \pm 0.35)M_v.
\end{equation}
Since the samples used for eq. (11) are
not statistically as uniform as simulation sample, one cannot take a detailed
comparison of observed results with simulations. Nevertheless, the
individual-ensemble duality of quasi-virialization gives a reasonable
explanation for the eq.(11) which used the sample consisting of
pre-dynamically-relaxed clusters.

Similar to eq.(11), the mass $M_x$ estimated from hydrodynamical equilibrium
model of X-ray gas is found to be, {\it on average}, in agreement with the
gravitational lensing-derived cluster masses $M_{lens}$. The correlation
between $M_X$ and $M_{len}$ in the scale range of  $0.25- 1.5$ $h^{-1}$ Mpc
is (Wu \& Fang 1997; Wu et al. 1998)
%eq11
\begin{equation}
M_{len}=(0.97\pm 0.44)M_X.
\end{equation}
This result also shows the feature of quasi-virialization: the hydrodynamical
equilibrium model-derived $M_x$ is reliable, while individual clusters may
deviate from virialization.

Moreover, the most recent statistics using the largest sample
of 149 clusters gives (Wu, Fang \& Xu 1998)
%eq13
\begin{equation}
\beta \equiv \frac{\sigma_v^2}{3kT/\mu m_p}=1.00 \pm 0.49,
\end{equation}
where the parameter $\mu = 0.59 $ is the mean molecular weight for
intracluster gas and $m_p$ is the proton mass. Eq.(13) means that the
temperature $T$ of X-ray emitting intracluster gas statistically can be
used as a virial indicator of the underlying gravitational potentials
of clusters, and is related to the velocity dispersion by virial relation as
%eq14
\begin{equation}
\frac{3}{2}kT=\frac{1}{2}\mu m_p \sigma_v^2.
\end{equation}
The current observations have revealed a constant temperature profile
within virial radius (Irwin, Bergman \& Evrard 1999).
Since the profile of $\bar{\sigma}_v$ is flat within radius $r<$ 6 $h^{-1}$ Mpc
(Fig. 10), eq.(14) indicates that the averaged temperature of clusters is
also approximately constant within the range of quasi-virialization.

\section{Temperature function of clusters}

\subsection{Temperature functions and dynamical state}

For the virialized and quasi-virialized clusters, we can define the
temperature function (TF) of clusters, $n_T(>T, r_{cl})$, which is the number
density of $r_{cl}$-clusters with temperature larger than $T$.
With eq.(14), the TFs can be calculated from the simulated velocity
dispersion function (VDF) $n_v(>\sigma_v, r_{cl})$, which is the number
density of $r_{cl}$-clusters with velocity dispersion larger than $\sigma_v$.
Actually, eq.(14) has been widely employed in the calculation of TF from
VDF (e.g. Klypin \& Rhee, G. 1994; Jing \& Fang 1994). Yet, as we have
emphasized, we are interested in the effect of dynamical state on
the TFs.

Fig. 13 plots the differential TFs, $n(T, r_{cl})=dn(>T, r_{cl})/dT$, for
$r_{cl} =$ 0.75 to 12 $h^{-1}$Mpc clusters in the three dark matter
models. In fact, the VDF-TF transfer [eq.(14)] may not be correct for
clusters of $r_{cl} =$ 12 $h^{-1}$ Mpc, which is plotted  only for 
comparison. 

If the system is fully virialized on scales $r<r_{cl}$, the TFs should be
basically scale-independent in this scale range, because virial temperature
$T$ is independent on radius, and clusters identified by $(T, r_{cl})$ is
the same as $(T,r'_{cl})$. Fig. 13 shows the significant scale-dependence
of the TFs, even at $r_{cl}=0.75$ $h^{-1}$ Mpc. Therefore, these systems
deviate from full virialization.

However, Fig. 13 shows that all the TFs of $r_{cl} \leq 6$ $h^{-1}$ Mpc have
similar shape, i.e. the TFs are scaling. This feature is from 
quasi-virialization. The TFs of $r_{cl} \geq 12h^{-1}$ Mpc clusters
do not join the scaling of $r_{cl} \leq 6$ $h^{-1}$ Mpc, and therefore,
it is not in quasi-virialization state.
This is expected, as Fig. 7 has shown that $r_{cl} \geq 12\ h^{-1}$ Mpc
clusters are different from that of $\leq 6$ $h^{-1}$ Mpc.

\subsection{The size and virialization of X-ray clusters}

As an application of simulated TF $n(T, r_{cl})$ (Fig. 13), we analyze the
problem of the size of X-ray clusters. Observationally, X-ray clusters are
described by their X-ray luminosity, flux weighted temperature and
morphology. Although
for some individual clusters we have known their X-ray isophotes and 
therefore the aperture of the X-ray emission, the X-ray luminosity is
generally published as pseudo-total luminosity which does not contain
information about the spatial scale of clusters. The Abell radius $r_{ab}$
is not used as a condition in X-ray cluster identification. Even for Abell
clusters, there is also the ambiguity of whether the X-ray luminosity comes
from cluster halos of larger than radius $r_{ab}=$ 1.5 $h^{-1}$ Mpc.
If clusters are fully virialized, the size uncertainty of the TFs is not
important since the TFs are size-independent.

However, in quasi-virialization state, the size-dependence of TFs is
substantial. For instance, we ploted the observed TF given by Edge
et al. (1990) and Henry \& Arnaud (1991) in Fig. 13. The LCDM and OCDM TFs
are in good agreement with the observed data only if $r_{cl}$ is in the
range 0.75 - 1.5 $h^{-1}$ Mpc. In other words, the LCDM and OCDM can pass
the TF test if most X-ray clusters used for the observed TF are on the
scales of 0.75 and 1.5 $h^{-1}$ Mpc. On the other hand, to fit the SCDM TF
with observed data, the size of X-ray clusters should be in the range
1.5 -3 $h^{-1}$ Mpc, which is larger than the results of LCDM and OCDM by
a factor of 2.

 The role of the size of X-ray clusters can be more clearly illustrated
by SCDM model with normalization amplitude $\sigma_8 =1$. Fig.14 shows the
TFs of the $\sigma_8 =1$ SCDM model. In this case, the model-predicted TF
can fit the data only if the size of X-ray clusters is in the range 6 to
12 $h^{-1}$ Mpc, which seems very unlikely. 

Thus, one can conclude that for either the model of SCDM or LCDM and OCDM, the
simulated TFs can fit with observations, if only the size of the X-ray
clusters used for constructing the TFs is, on average, larger than their
virial radius. This is consistent with the fact that the configuration of
X-ray clusters generally is irregular and substructured.

Obviously, the scale-dependence of the TFs can be employed for discriminating 
among
models if the information of the size of X-ray clusters is available.
Currently, we are lacking of the data of X-ray cluster sizes. Instead, one
can consider the test of TF plus two-point correlation function. As it
has been shown in Figs. 13 and 14, the observed TFs can set a lower limit
to the scale of clusters. For instance, the data of Edge et al. (1990)
require that $r_{cl}$ should be $\geq 0.75$ $h^{-1}$ for the LCDM
and OCDM, $\geq 3$ $h^{-1}$ Mpc for the SCDM, and $ \geq 6$ $h^{-1}$ Mpc
for the $\sigma_8=1$ SCDM. On the other hand, for a given abundance, the
larger the scale $r_{cl}$, the smaller the correlation length
(Xu, Fang \& Deng 1999). 
At the abundance $d > 70$ $h^{-1}$ Mpc,
the correlation length is found to be $r_0 > 15 h^{-1}$ Mpc if only 
$r_{cl} \leq 6$ $h^{-1}$ Mpc in the LCDM and OCDM; and 
$r_{cl} \leq 3$ $h^{-1}$ in the SCDM.

Since the observed correlation length $r_0$ of X-ray clusters is 
$r_0 > 15 h^{-1}$ Mpc. The LCDM and OCDM are able to pass the two constraints
to $r_{cl}$ simultaneously: 1.) TF requires $r_{cl}\geq 0.75$ $h^{-1}$ Mpc,
and 2.) correlation length requires $r_{cl} \leq 6$ $h^{-1}$ Mpc. Yet, the
SCDM with $\sigma_8=1$ is clearly in trouble as 1.) TF requires $r_{cl}\geq 3$
$h^{-1}$ Mpc, and 2.) correlation length requires at least
$r_{cl} \leq 3$ $h^{-1}$ Mpc.

\section{Conclusions}

Using N-body simulated samples in models of the SCDM, LCDM and OCDM, we
have analyzed the dynamical state of the DWT multi-scale identified clusters.
We showed that state of quasi-virialization exists in the dynamical
evolution of these clusters, i.e. while the internal structure of individual
clusters significantly differs from dynamical relaxation on scales larger
than virial radius $r_{vir}$, some statistical features of the clusters are
approximately the same as those for the virialized systems. Actually, the
virial statistical features, like
$\sigma_v \propto M^{1/2}$, $\xi_{cm}(r) \propto r^{-2.5}$, are scaling
relations. 
The dynamical reason of the existence of quasi-virialization is 
that some of the scaling properties of dynamical relaxed systems of cosmic 
gravitational clustering  approximately hold beyond the fully virialization 
regime. The scaling behavior of mass functions of the quasi-virialized
clusters can also be repeated by semi-analytic calculation on
collapsed and uncollapsed halos in Press-Schechter formalism. 

Consequently, in terms of statistical description, the cluster galaxies and
X-ray gas can be used as virial tracers of cluster potential on scales beyond
the full virialization. This result provides a good explanation that the
virial masses and X-ray masses are basically the same as the gravitational
lensing determined masses on scales beyond the full virialization. It also
explained the tight correlation between the velocity dispersion of optical
galaxies, and the temperature of the clusters. The virial mass estimator
based on the assumptions of isothermal and hydrostatic model is applicable
on scales as large as about $r_{cl}=6h^{-1}$Mpc.

In the quasi-virialization state, the TFs have scaling. It is potentially
important for model-discrimination. A very preliminary result given by the
test of the scale-dependence of the TFs showed that the LCDM and OCDM are
favored.

\acknowledgments

Wen Xu thanks the World Laboratory for a scholarship. This project
was done when W.X. was visiting at Physics Department,
University of Arizona.

\clearpage

\begin{deluxetable}{cccccccc}
\tablecaption{Mean number of $\alpha \equiv d\ln M/d\ln r$ in 
  the range from 0.5 to 24 $h^{-1}$Mpc }
%\tablewidth{0pt}
\tablewidth{38pc}
\tablehead{
 \colhead{$r_{cl}$} & \colhead{0.75}  & \colhead{1.5}
 & \colhead{3.} & \colhead{6.} & \colhead{12.} & \colhead{24.}
}
\startdata
OCDM &0.50$\pm$0.06 &0.49$\pm$0.06 &0.52$\pm$0.05&0.53$\pm$0.05&
 0.68$\pm$0.04 & 0.67$\pm$0.06   \nl
LCDM &0.51$\pm$0.06 &0.47$\pm$0.06 &0.51$\pm$0.05&0.55$\pm$0.04&
 0.67$\pm$0.04 & 0.65$\pm$0.06   \nl
SCDM &0.40$\pm$0.08 &0.48$\pm$0.06 &0.44$\pm$0.06&0.49$\pm$0.07
& 0.67$\pm$0.06 & 0.75$\pm$0.07  \nl
\enddata
 \end{deluxetable}

\clearpage

\figcaption{The mass function of clusters with $r_{cl}=1.5, 3, 6, 12, 24 
h^{-1}$ Mpc of N-body simulation (circle) and of the PS formalism 
predictions (solid line) for SCDM. 
For the PS results, the MF of $r_{cl}=1.5 h^{-1}$
Mpc is the number density of collapsed halos, and the MFs of
$r_{cl}=3, 6, 12, 24 h^{-1}$ Mpc are the number density of 
collapsed and uncollapsed $r_{cl}$-halos.
\label{fig1}}

\figcaption{Logarithmic mass contrast as a function of $r_{cl}$ for 
$r_{cl}$-clusters with richness $d=$ 30, 50 and 100 $h^{-1}$Mpc.
\label{fig2}}

\figcaption{A projected distribution of $r_{cl}$-clusters identified
from one realization of OCDM N-body simulation within box 310 $h^{-3}$
Mpc$^3$. It shows the top 25 of massive clusters for $r_{cl}=$ 0.75, 1.5,
3, 6, 12 and 24 $h^{-1}$ Mpc. The legends of the symbols are: plus for
0.75, diamond for 1.5, circle for 3, triangle for 6, square for 12 and
giant circle for 24 $h^{-1}$ Mpc. The linear size of the symbols is roughly
equal to the corresponding $r_{cl}$, but slightly amplified for smaller 
$r_{cl}$ for clarity.
\label{fig3}}

\figcaption{Average number of substructures per $r_{cl}$-cluster as a 
function of logarithmic counts of particles in each cluster.
The data is from one realization of the OCDM model simulation.
\label{fig4}} 

\figcaption{Probability distributions of number of substructures 
per cluster for top 59 richest $r_{cl}=24h^{-1}$Mpc clusters in a 
310 $h^{-1}$ Mpc simulation box. Results of three cosmological models
are shown.
\label{fig5}}

\figcaption{The mean number of substructures per cluster as a function
of $r_{cl}$ for top 59 richest clusters in a 310 $h^{-1}$ Mpc simulation 
box for models of the SCDM, LCDM and OCDM. Error bars are the standard
deviation in $N$. The results of SCDM and OCDM at redshift $z=0.5$ are 
also shown, but without error bars for clarity. To avoid overlapping with 
the line of SCDM at $z=0$, the line for LCDM at $z=0.5$ is not shown.
\label{fig6}}

\figcaption{Three dimensional velocity dispersion $\sigma_v$ vs. 
mass of clusters of $r_{cl}=$ 0.75, 1.5, 3, 6, 12, and 24 $h^{-1}$ Mpc. 
The data are from three realizations of OCDM model at $z=0$. The mean 
logarithmic velocity dispersion, $\overline{\log \sigma}_v$, within each 
bin of width 0.1 dex in $\log M(r_{cl})$ are shown by thick crosses, for 
which the horizontal bars represent the bin widths, and the vertical 
error bars are the standard deviation of $\log \sigma_v$ within each bin.
\label{fig7}}

\figcaption{$\gamma$ vs. richness ($d$) of clusters of
$r_{cl}=$ 0.75, 1.5, 3, 6, 12, and 24 $h^{-1}$ Mpc. The results for three 
models SCDM, LCDM and OCDM at redshifts $z=0$ and $z=0.5$ are given.
Error bars are only illustrated for $r_{cl}=12$ $h^{-1}$ Mpc.
Errors for other $r_{cl}$ are similar to $r_{cl}=12$ $h^{-1}$ Mpc,
but having not been illustrated for clarity.
\label{fig8}}

\figcaption{The redshift evolution of $\gamma$ for the top 59 clusters of
$r_{cl}=$ 1.5, 12, and 24 $h^{-1}$ Mpc in models SCDM , LCDM and OCDM.
Error bars are illustrated only for $r_{cl}=12 h^{-1}$ Mpc.
\label{fig9}}

\figcaption{The averaged profiles of velocity dispersion within radius $r$
for top 50 massive clusters in a 310 $h^{-1}$Mpc simulation
box. Clusters of $r_{cl}=$ 0.75, 1.5, 3, 6, 12, and 24 
$h^{-1}$ Mpc of the SCDM and OCDM are shown.
\label{fig10}}

\figcaption{The cross-correlation function between cluster and its particles
for clusters with $r_{cl}$= 1.5, 6 and 24 $h^{-1}$ Mpc. The 
richness for all $r_{cl}$ clusters is taken to be $d= 90$ $h^{-1}$ Mpc. 
The heavy and thin lines are for OCDM and SCDM, respectively.
\label{fig11}}

\figcaption{The cross-correlation function between cluster and its particles
for $r_{cl}=$ 6 $h^{-1}$ Mpc clusters with richness  
$d=$ 40 - 120 $h^{-1}$ Mpc in the OCDM model.
\label{fig12}}

\figcaption{Temperature function $n(T, r_{cl})$ averaged from 8 realizations
for clusters of $r_{cl}=$ 0.75, 1.5, 3, 6 and 12 $h^{-1}$ Mpc in the SCDM, 
LCDM and OCDM models. The TFs of $r_{cl}=12$ $h^{-1}$Mpc (dashed lines) are 
clearly not a member of the scaling family. The observations of Henry \&
Arnaud (1991) and  Edge et al. (1990) are illustrated as triangles and 
circles, respectively.
\label{fig13}}

\figcaption{Temperature functions averaged from 3 realizations for clusters 
of $r_{cl}=$ 0.75, 1.5, 3, 6 and 12 $h^{-1}$ Mpc in the SCDM model with 
COBE normalization ($\sigma_8=1.0$). The observational points are the same 
as in Fig. 13.
\label{fig14}}

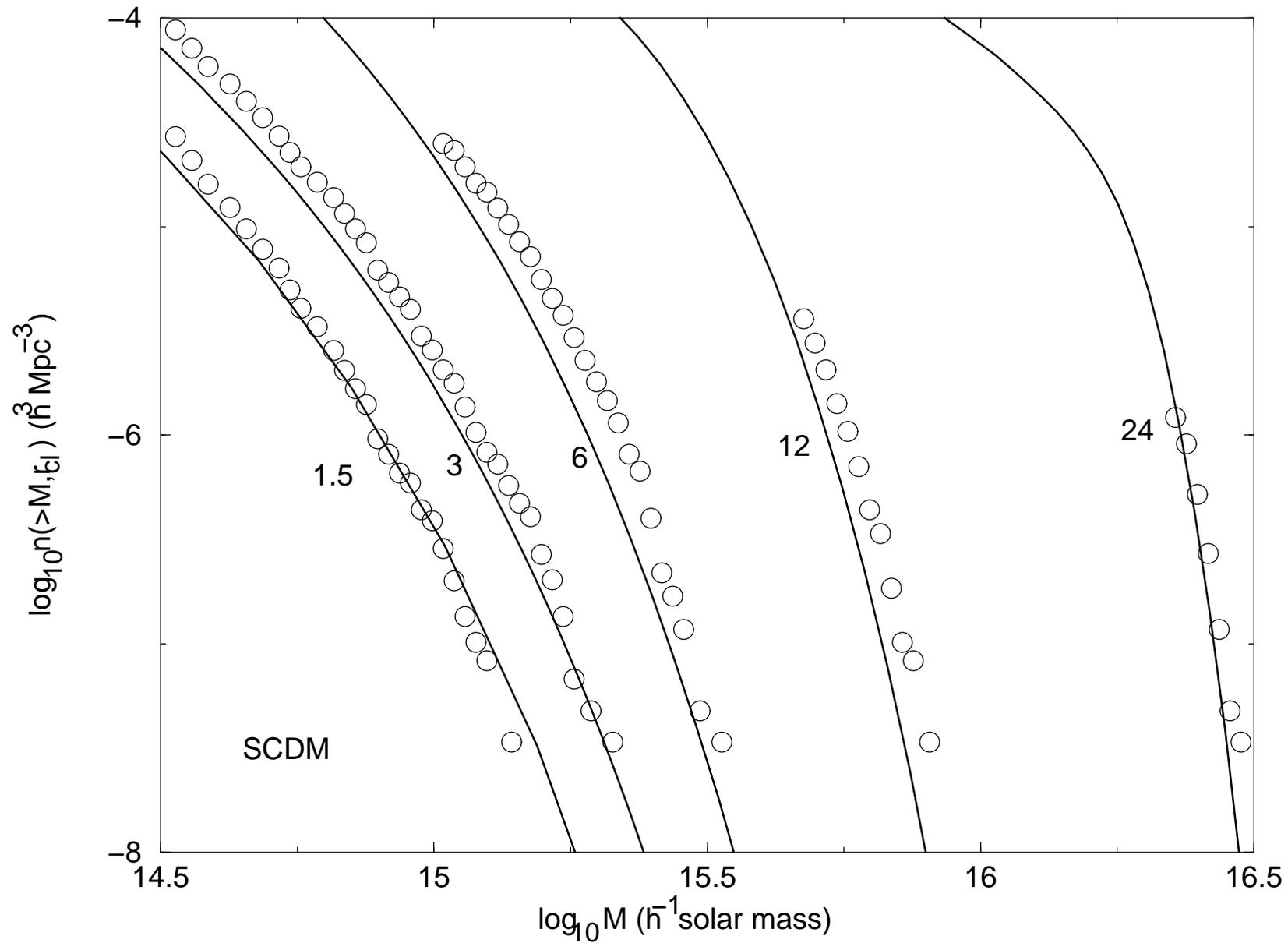
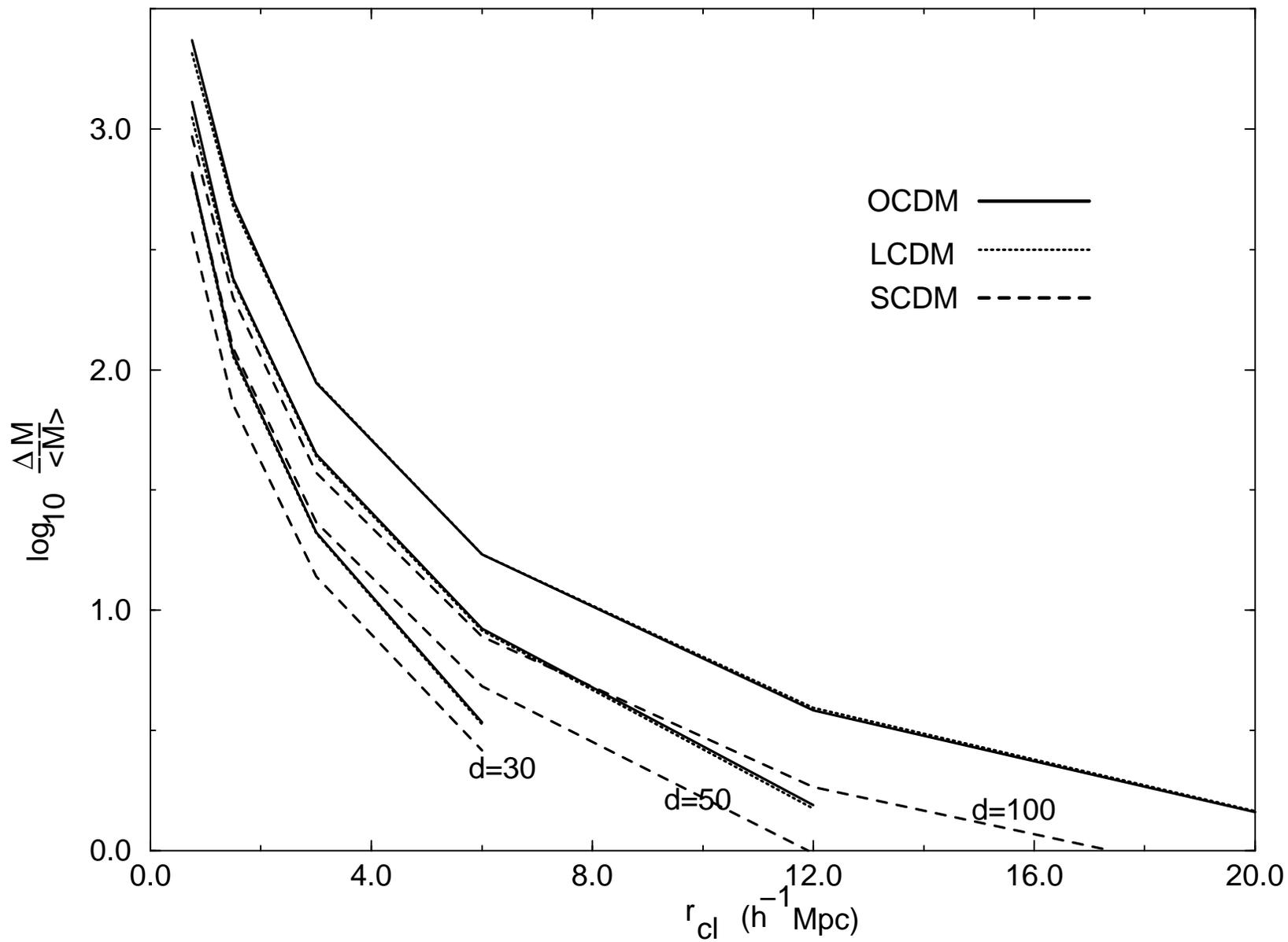
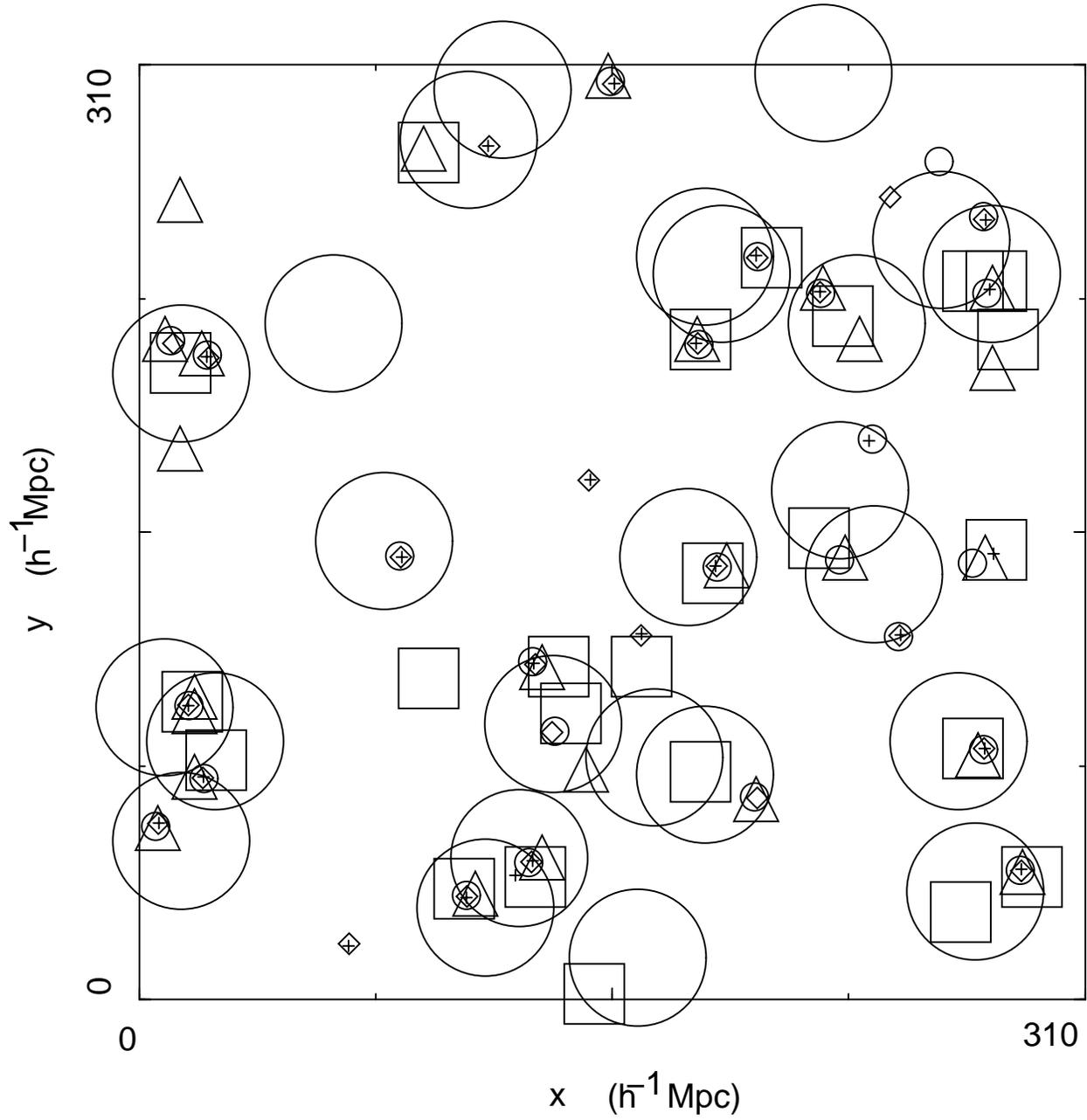
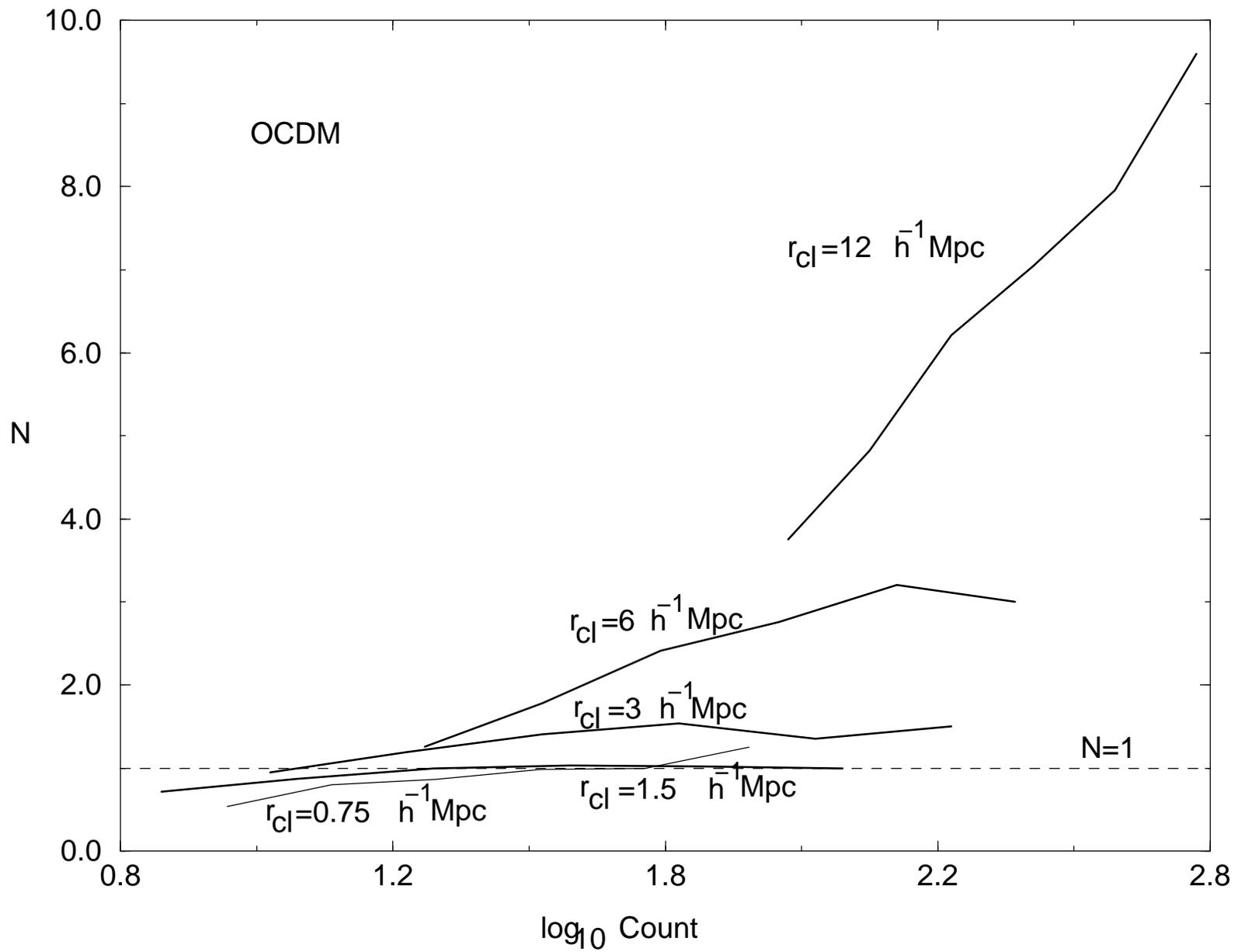
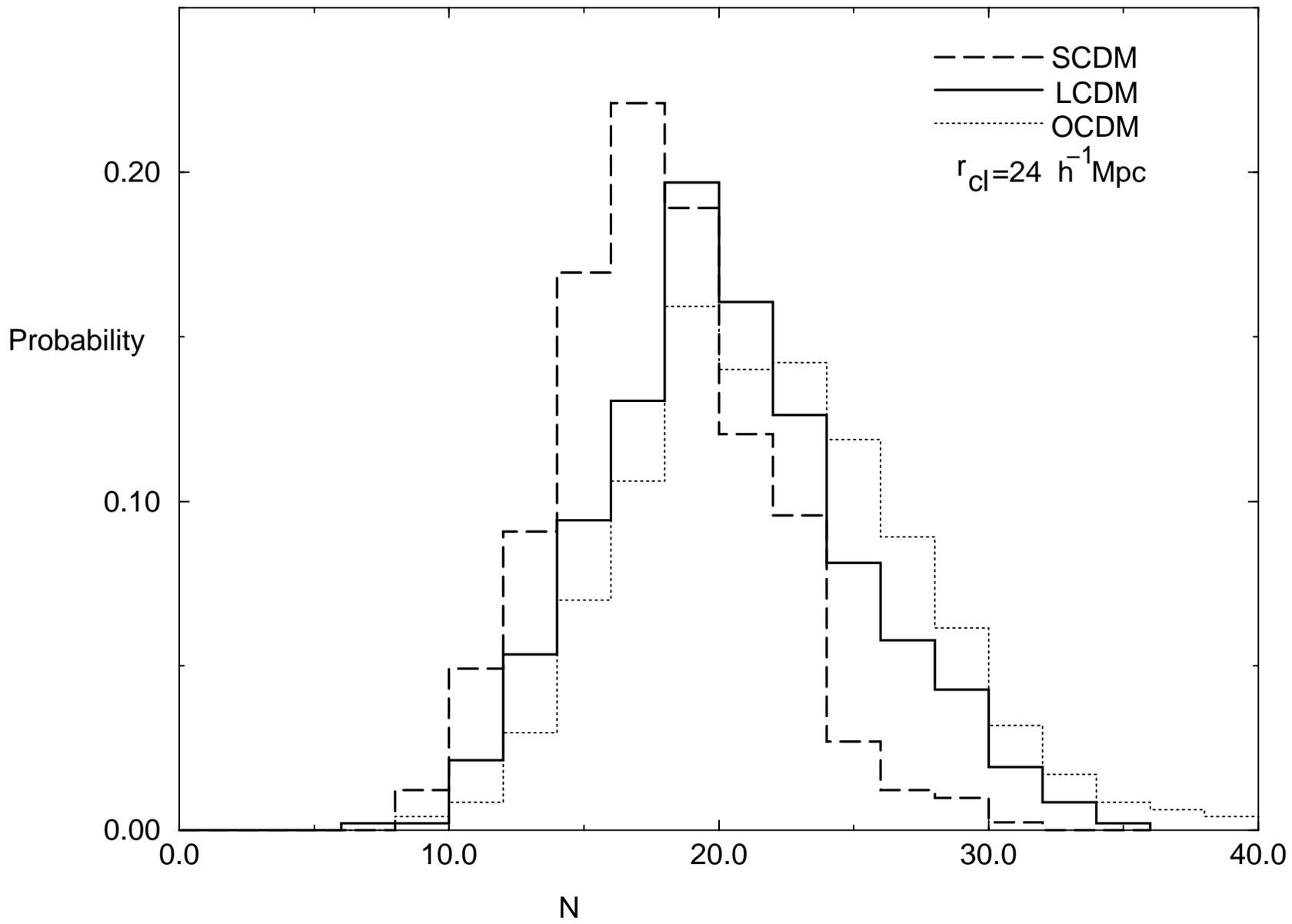
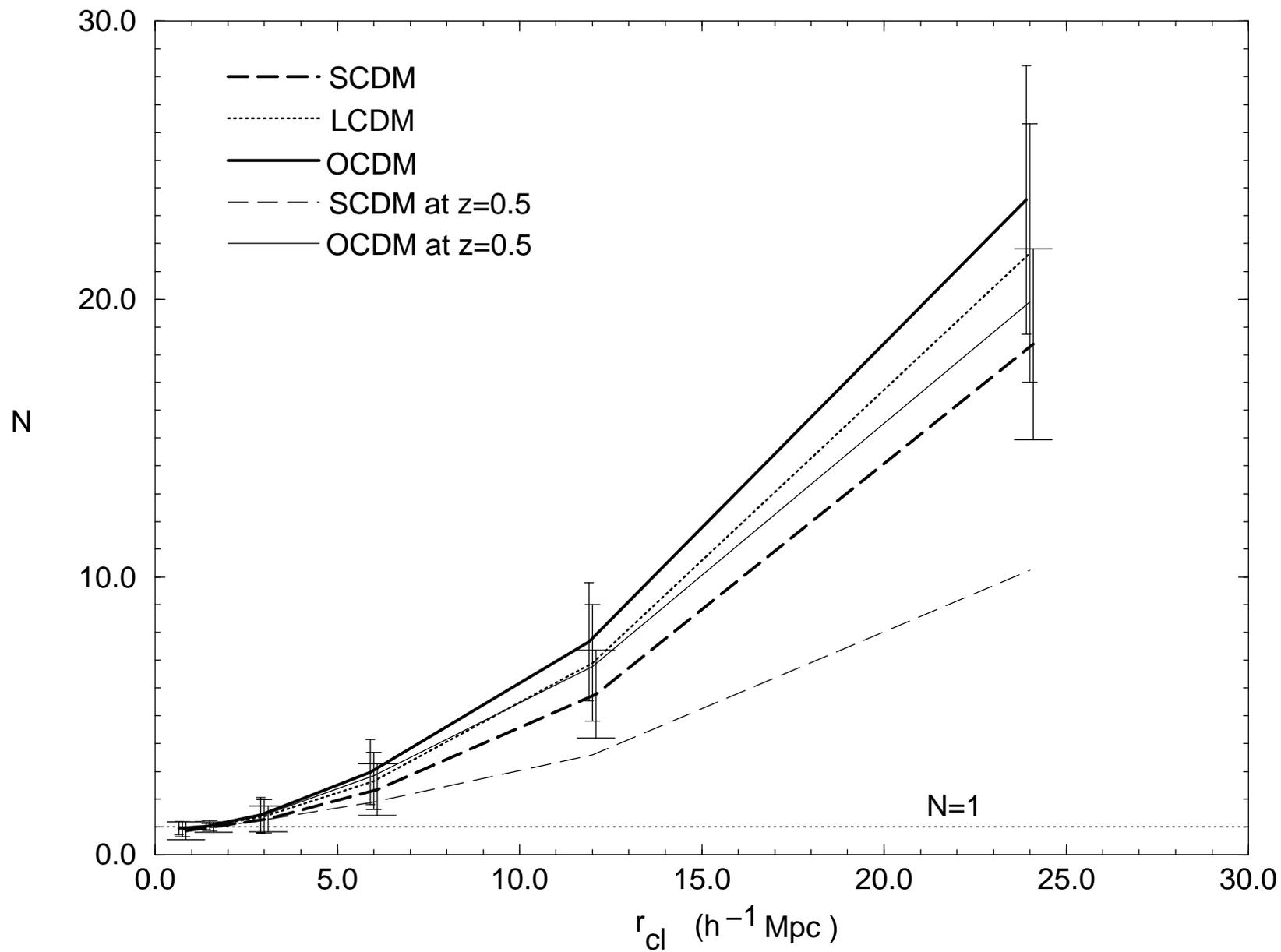
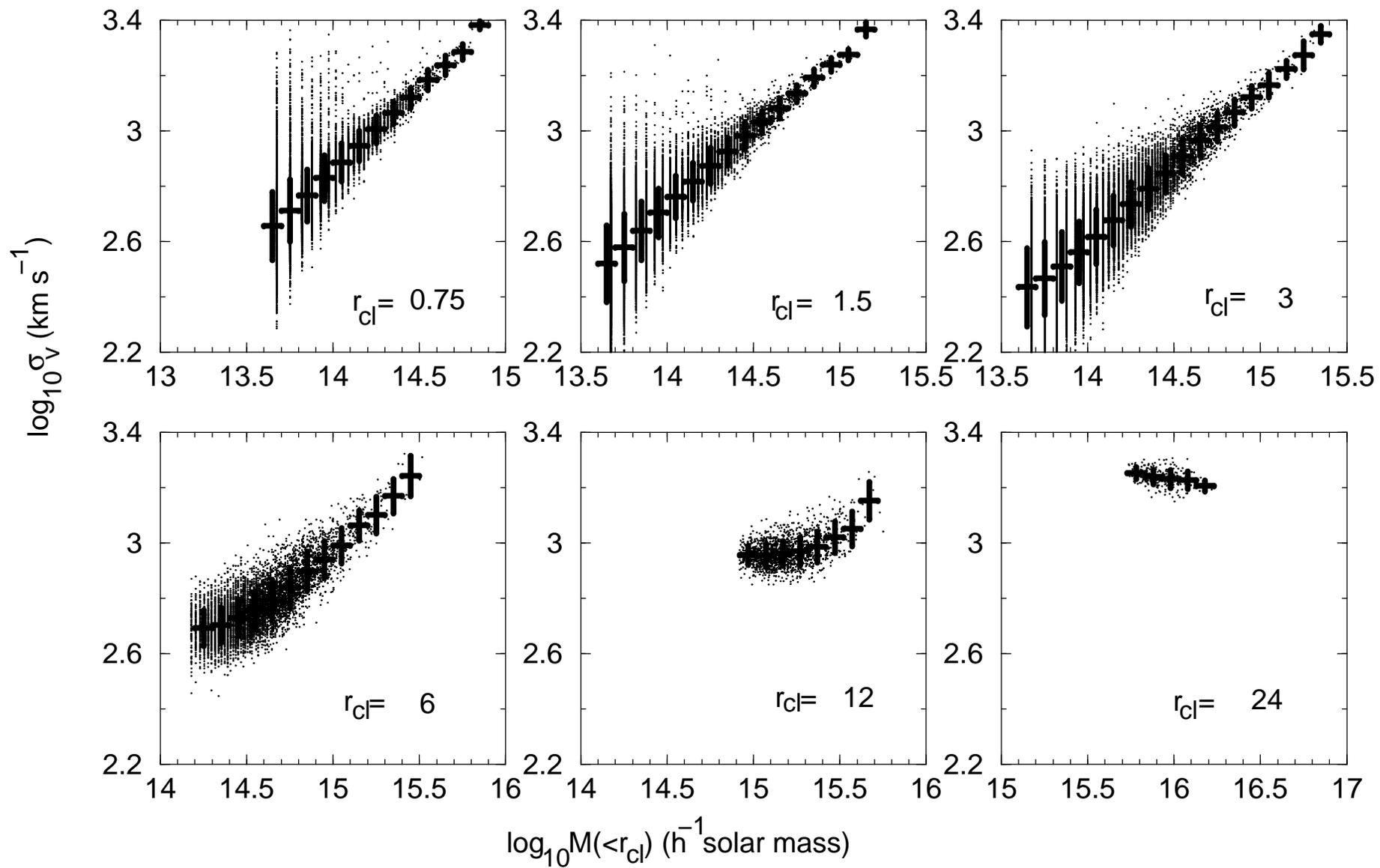
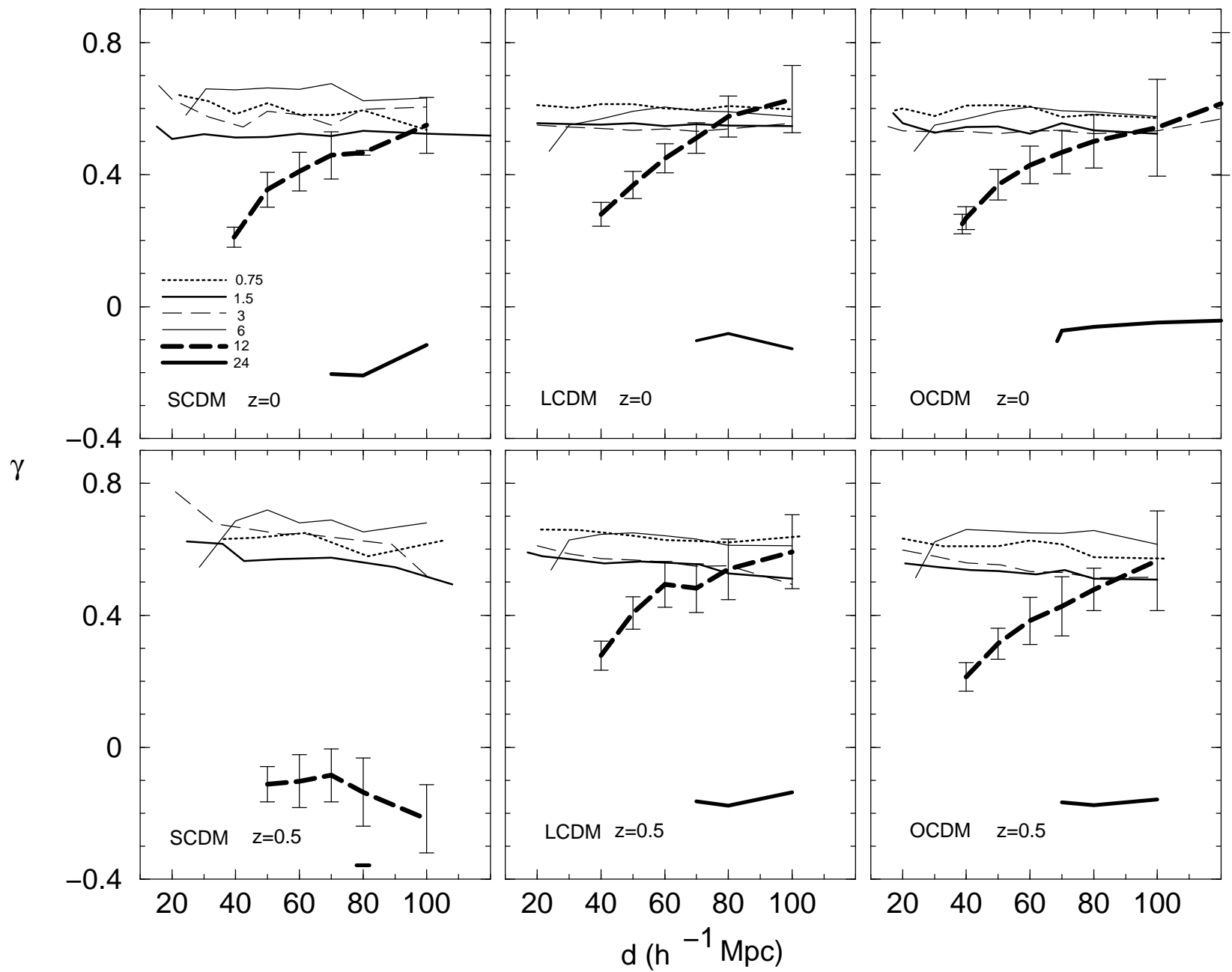
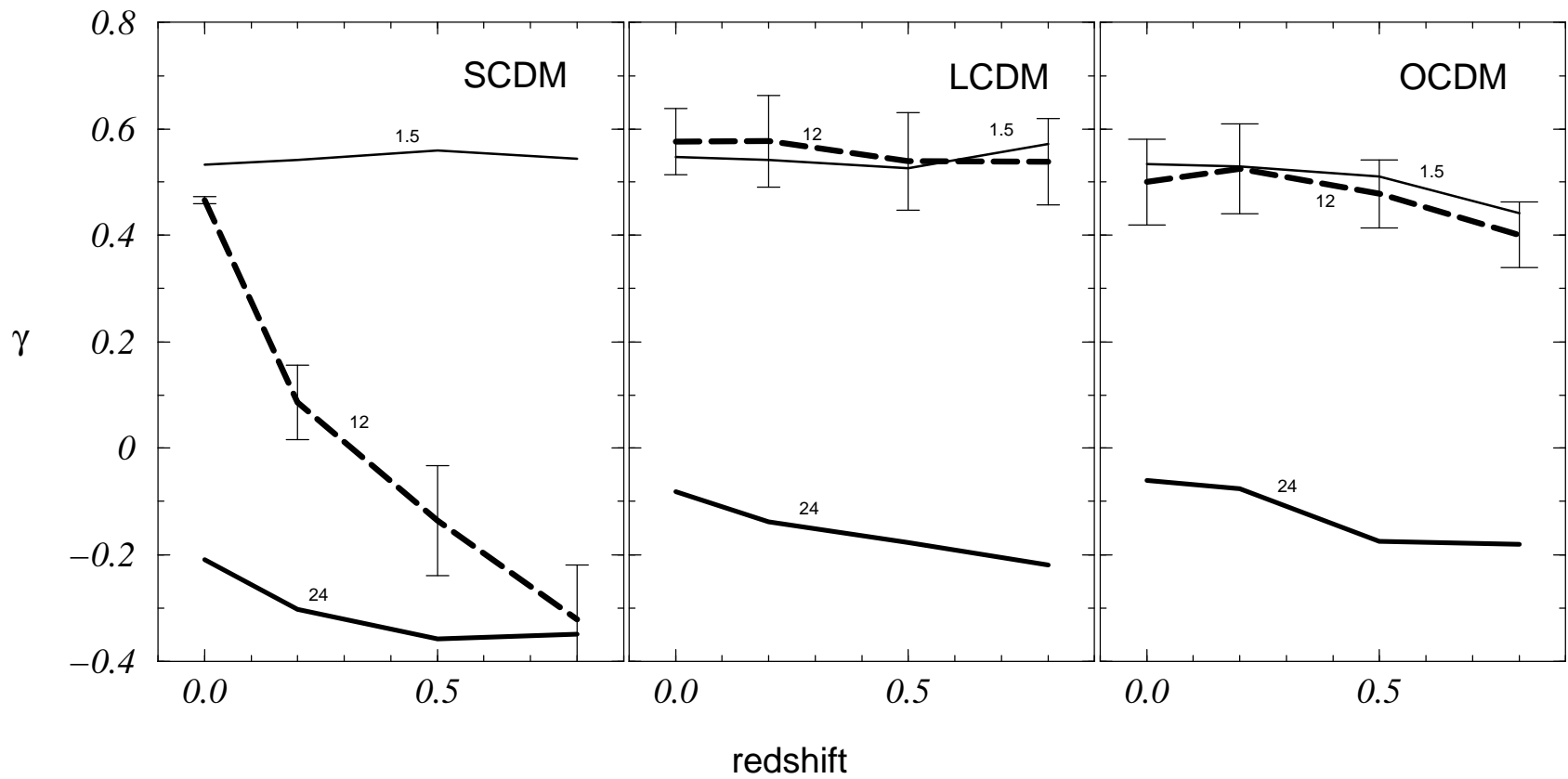
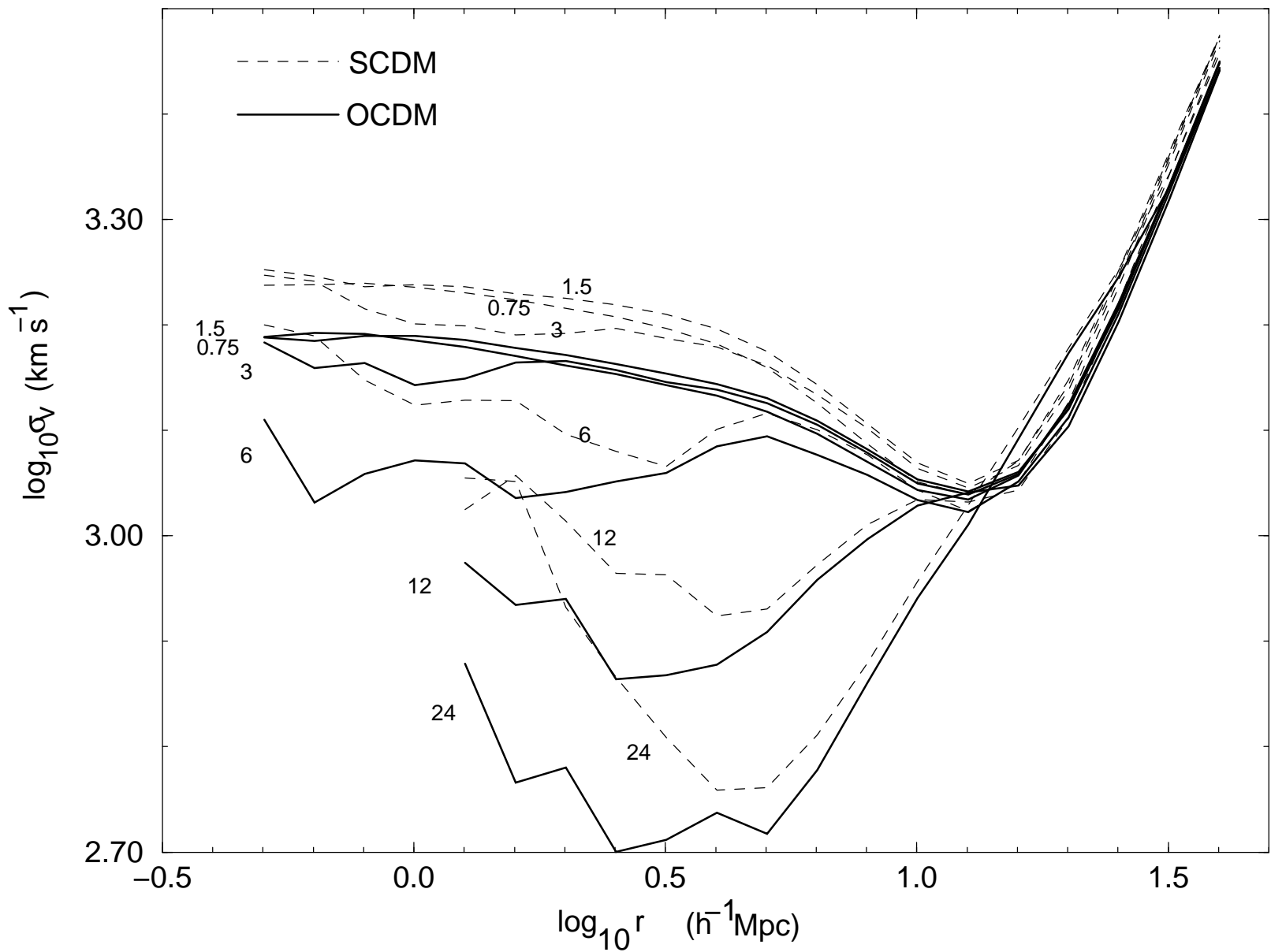
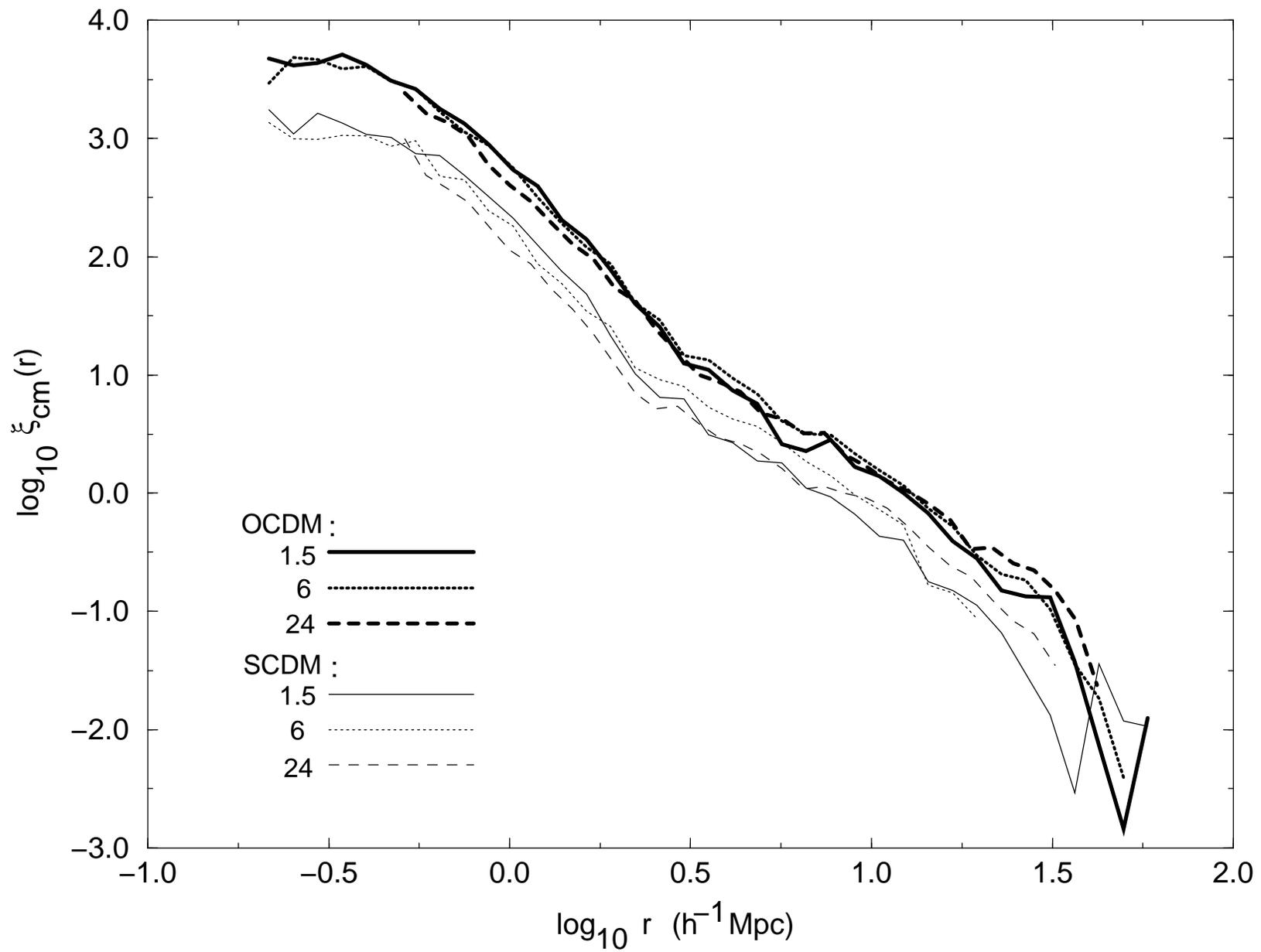
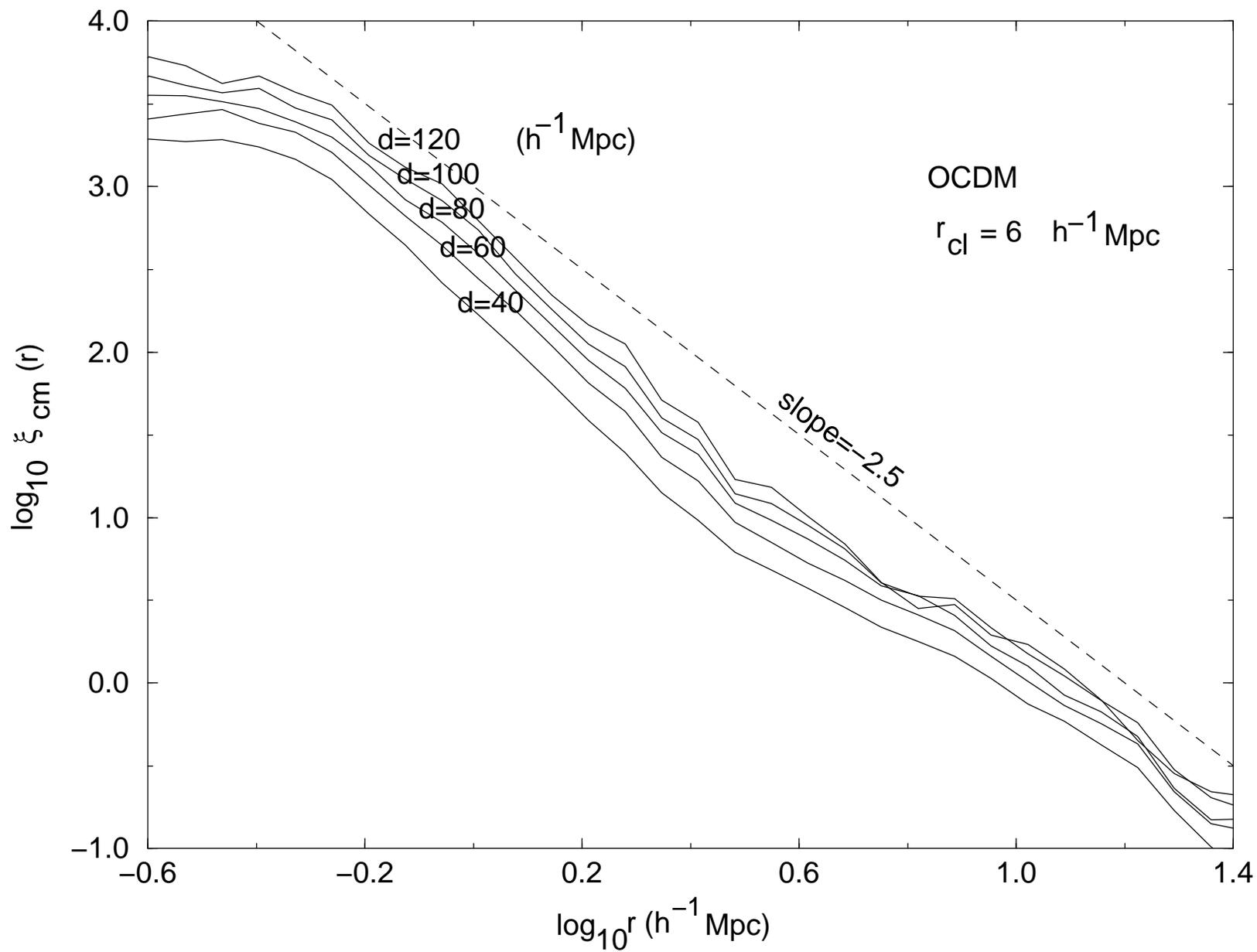
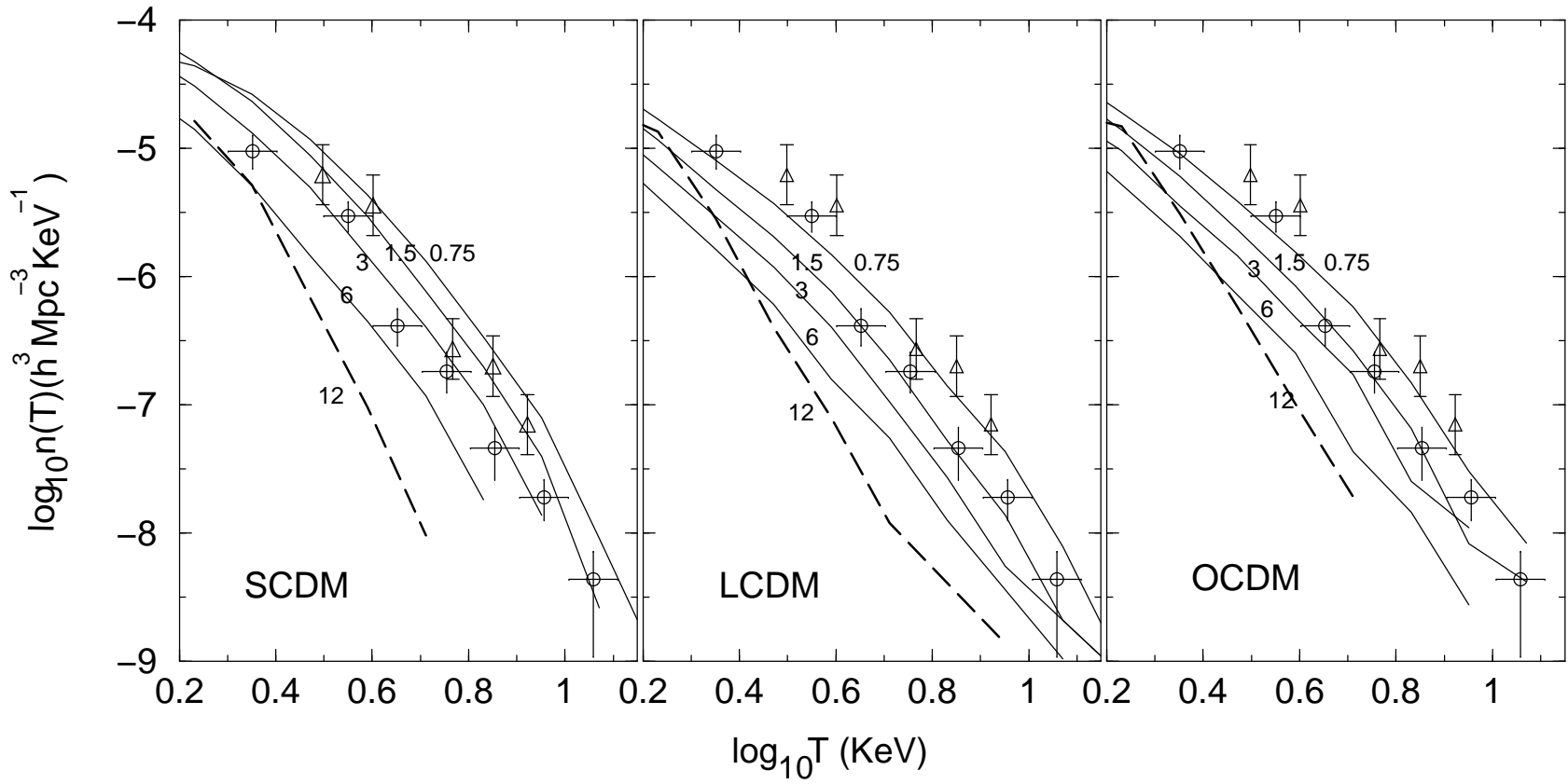
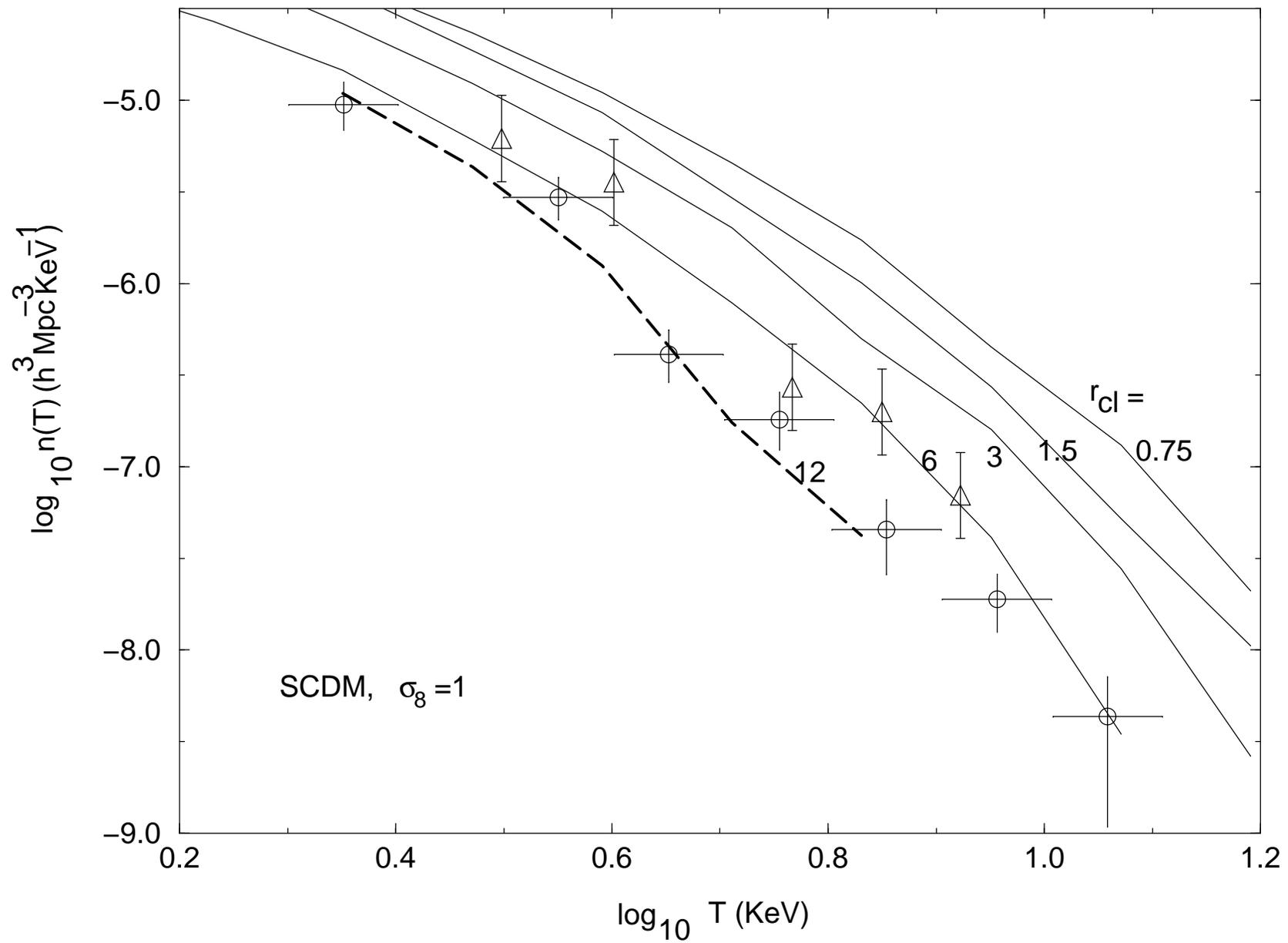
\end{document}